\documentclass[journal=jacsat,manuscript=article]{achemso}
\usepackage[utf8]{inputenc}
\usepackage[T1]{fontenc}
\usepackage{graphicx}
\usepackage{babel}
\usepackage{hyperref}
\makeatletter
\usepackage{gensymb}
\usepackage{geometry}
\usepackage{array}
\usepackage{enumitem}
\usepackage{xurl}
\usepackage{diagbox}
\usepackage[table,xcdraw]{xcolor}
\usepackage{pifont}
\usepackage{amssymb}
\usepackage{subfiles}
\usepackage{caption}
\DeclareUnicodeCharacter{2212}{-}
\title{Gas permeability, diffusivity, and solubility in polymers: Simulation-experiment data fusion and multi-task machine learning}
\author{Brandon K. Phan}
\affiliation{School of Materials Science and Engineering, Georgia Institute of Technology, Atlanta, Georgia 30332, United States}

\author{Kuan-Hsuan Shen}
\affiliation{School of Materials Science and Engineering, Georgia Institute of Technology, Atlanta, Georgia 30332, United States}

\author{Rishi Gurnani}
\affiliation{School of Materials Science and Engineering, Georgia Institute of Technology, Atlanta, Georgia 30332, United States}

\author{Huan Tran}
\affiliation{School of Materials Science and Engineering, Georgia Institute of Technology, Atlanta, Georgia 30332, United States}

\author{Ryan Lively}
\affiliation{School of Chemical and Biomolecular Engineering, Georgia Institute of Technology, Atlanta, Georgia 30332, United States}

\author{Rampi Ramprasad}
\affiliation{School of Materials Science and Engineering, Georgia Institute of Technology, Atlanta, Georgia 30332, United States}

\makeatletter
\email{rampi.ramprasad@mse.gatech.edu}
\makeatother
\begin{document}

\maketitle

\newpage

\begin{abstract}

Machine learning (ML) models for predicting gas permeability through polymers have traditionally relied on experimental data. While these models exhibit robustness within familiar chemical domains, reliability wanes when applied to new spaces. To address this challenge, we present a multi-tiered multi-task learning framework empowered with advanced machine-crafted polymer fingerprinting algorithms and data fusion techniques. This framework combines scarce "high-fidelity" experimental data with abundant diverse "low-fidelity" simulation or synthetic data, resulting in predictive models that display a high level of generalizability across novel chemical spaces. Additionally, this multi-task scheme capitalizes on known physics and interrelated properties, such as gas diffusivity and solubility, both of which are closely tied to permeability. By amalgamating high-throughput generated simulation data with available experimental data for gas permeability, diffusivity, and solubility for various gases, we construct multi-task deep learning models. These models can simultaneously predict all three properties for all gases under consideration. With markedly enhanced predictive accuracy, particularly compared to traditional models reliant solely on experimental data for a singular property. This strategy underscores the potential of coupling high-throughput classical simulations with data fusion methodologies to yield state-of-the-art property predictors, especially when experimental data for targeted properties is scarce.
\end{abstract}

\newpage
% \linenumbers

\section{Introduction}

Polymer-based gas and solvent separation membrane technologies have significantly impacted a diverse range of applications, including carbon capture, water purification, drug delivery, and food packaging.\cite{Ferreira2016Polysaccharide,Baker_2001Membrane} Crucial to propelling widespread adoption and advancement of this technology is the identification and design of polymer materials endowed with a desired set of properties and performance attributes. A key figure of merit in gas separations is gas permeability, which describes the movement of gas molecules into and through a polymer material. Based on the solution-diffusion model\cite{Wijmans_Baker_1995}, gas permeability ($P$) through a membrane is defined as the product of gas diffusivity ($D$) and gas solubility ($S$):
\begin{equation}
P=DS\label{eq:pds}
\end{equation}
Capabilities that can accurately and rapidly predict gas permeability across a diverse range of gases and polymer chemistries can be transformational and facilitate the discovery and development of new sustainable high-performance polymer membranes.\cite{tran2023informatics2, barnett2020designing}

Traditionally, the measurement of gas permeability relies on the constant volume permeation technique\cite{moore2004characterization}, which, though serving as the primary benchmark, is both time and resource-intensive. In search of alternative approaches, classical molecular dynamics (MD) simulations have emerged as a complementary pathway to estimate gas permeability.\cite{Müller1994} However, the fidelity of these simulations is constrained by the intrinsic limitations of the classical force fields employed and timescales that are computationally accessible. As a result, they can only achieve, at best, semi-quantitative agreement with experimental measurements, despite correctly capturing general trends.

In recent times, data-driven machine learning (ML) methods have achieved remarkable strides, fundamentally reshaping the landscape of materials property predictions and the tailored design of materials with specific target characteristics.\cite{audus2017polymer, batra2021emerging, chen2021polymer,tran2023informatics2, zhu2020polymer, barnett2020designing,wu2022rational,chen2020frequency} ML methods have found extensive applications in the polymer gas transport domain, encompassing a diverse array of studies varying in the number of polymers investigated and the types of features used to train models. An early example of this is the work by Wessline et al. in 2006, where a neural network was used to correlate the infrared spectra of 33 polymers with their carbon dioxide permeability.\cite{wessling1994modelling} In a more current study, Yuan et al. utilized Multivariate Imputation by Chained Equations (MICE) to predict missing gas permeability values in a dataset spanning hundreds of polymers across six gases.\cite{yuan2021imputation} These examples only scratch the surface. In a comprehensive perspective paper, Ricci et al. delve deeper into the evolution of ML in modeling gas separation with polymer membranes, highlighting strategies, challenges, and future directions.\cite{ricci2023perspective}

These informatics approaches require a critical initial step: defining the feature space in which the models are trained by mapping features to the properties being learned. Early machine learning (ML) studies employed simple feature sets; for instance, in a 2006 study by Wang et al., six features related to the experimental setup, such as temperature, feed gas flux, and permeate-side pressure, were used.\cite{wang2006radial} These approaches have transitioned to incorporate more descriptive and comprehensive features, capturing atomistic to morphological structural details.\cite{rogers2010extended,landrum2013rdkit} In this paradigm, a polymer's chemical structure is converted into a machine-readable numerical representation, commonly known as a fingerprint or feature vector. This fingerprint allows an ML algorithm in the second step (during the training phase) to discern intricate chemistry-morphology-property relationships and subsequently generate predictive models for the properties. While traditional hand-crafted fingerprints\cite{huan2015accelerated,le2012quantitative} have conventionally represented polymer structures in machine learning models, recent endeavors have expanded the horizons of this methodology and have led to learned fingerprinting techniques, which we adopt in this study. These techniques involve machine learning key features directly from polymer repeat units, offering faster feature extraction with comparable accuracy.\cite{gurnani2022polymer,kuenneth2023polybert} Despite these advancements, a common challenge these methods encounter is extrapolating outside of the known polymer-property space, i.e., outside of the training data space. \cite{ramprasad2017machine} Exploring new chemical spaces through various avenues, including experiments, simulations, and machine learning models poses unique limitations that necessitate innovative solutions.

In the present contribution, we demonstrate the power of multi-task (MT) learning, harnessing both experimental and computational data to address and bridge the shortcomings outlined above, to build a best-in-class gas transport property predictor. MT learning is a type of transfer learning in which a model is trained on more than one task, learning multiple properties and/or data sources simultaneously.\cite{hutchinson2017overcoming} In contrast, single-task (ST) learning involves the consideration of a singular property and data source. The MT architecture, which integrates various data sources and exploits underlying correlations and calibrations, has shown improved predictive performance and enhanced transferability, compared to ST methods.\cite{caruana1998multitask,patra2020multi} In the polymer gas transport ML space, MT learning has been commonly implemented by incorporating permeability data for various gases and utilizing datasets that encompass a broad spectrum of properties, including mechanical, thermal, and thermodynamic.\cite{kuenneth2021polymer,yang2022machine,barnett2020designing} We expand on these previous works by utilizing MT learning in two novel ways. The first aspect leverages data fidelity by fusing "high-fidelity" experimental data with "low-fidelity" simulation data. While experimentally measured data serves as the ground truth, it often grapples with constraints stemming from labor-intensive protocols and associated expenses. Conversely, simulation-generated data can be produced on a grander scale, but it may exhibit diminished accuracy due to necessary approximations made in the theory to make the simulations practical. MT algorithms learn to calibrate the low-fidelity (simulation) data against the high-fidelity (measured) data across the whole space of the data, thus leading to a high level of generalizability.\cite{venkatram2020predicting,patra2020multi} Typically, gas simulations have been used to validate ML predictions. Here, we integrate the simulation data into the model itself. 

The second innovative aspect of the MT learning approach extends the general gas permeability ML model to include directly correlated gas transport data; diffusivity and solubility. Gas transport experiments reported in the literature do not always include all three properties, and they tend to focus on testing specific gases of interest. Consequently, some property values may not be available for certain cases. MT learning offers a solution to this challenge by drawing on available properties to learn correlations between them and make effective generalizations.\cite{kuenneth2021polymer} Incorporated within these two outlined MT aspects is the integration of gas transport data spanning a variety of gases. Our MT learning strategy leverages potential correlations between the transport characteristics of multiple similar (or dissimilar) gases.\cite{yuan2021imputation} A unified model that harnesses data from (1) diverse sources (i.e., measured and simulated), (2) spanning multiple correlated properties (i.e., $P$, $D$, \& $S$), and (3) for various gases, can lead to enhanced predictive performance and generalizability, as will be demonstrated here.

\begin{figure}[H]
\includegraphics[width=.9\columnwidth]{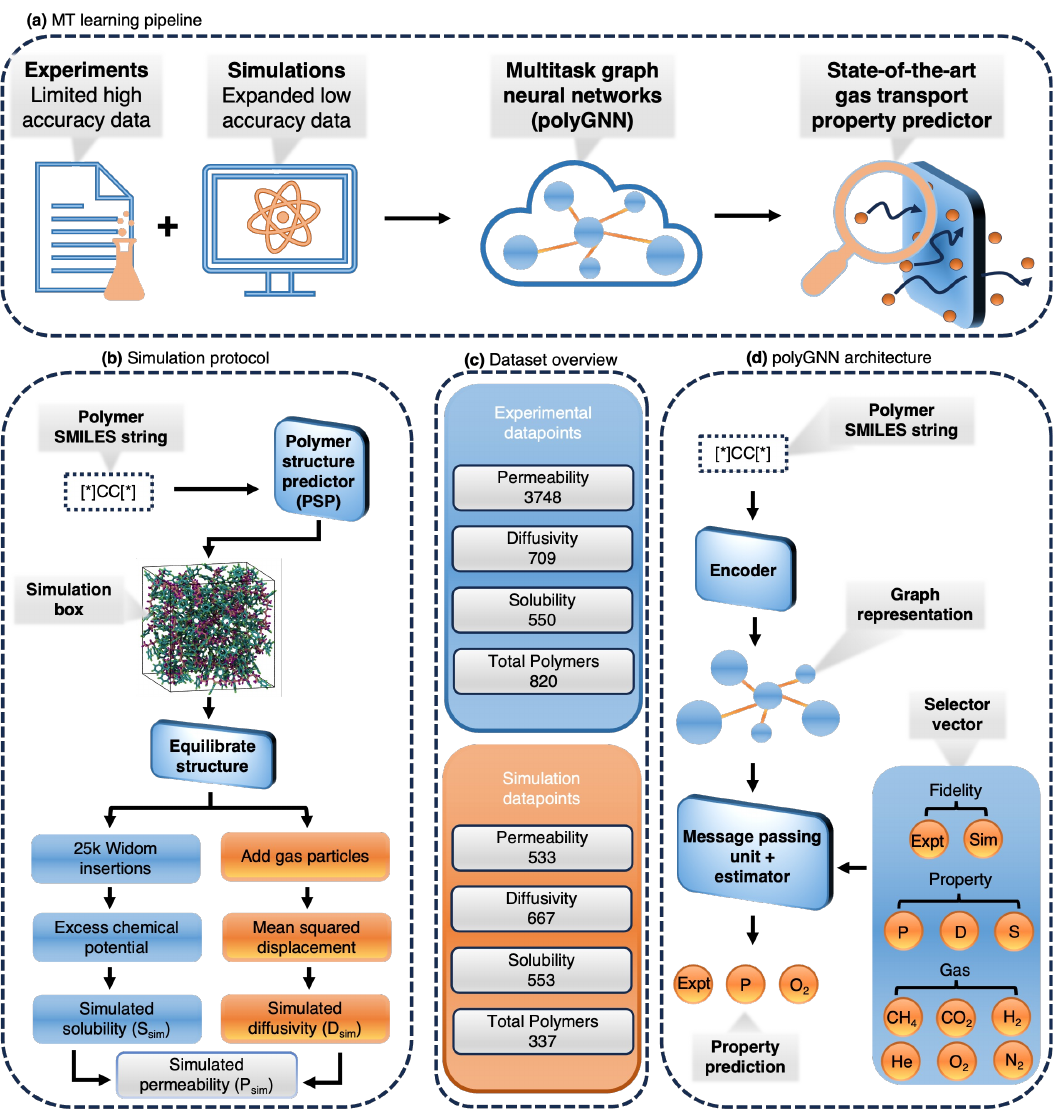}
\caption{
\textbf{(a)} \textbf{MT learning pipeline.} Our innovative multi-task learning approach employs the fusion of experimental and simulation data, harnessed through the power of polyGNN, a graph neural network architecture, to construct a state-of-the-art predictor for gas transport properties
\textbf{(b)}\textbf{ Simulation protocol.} The process begins with a polymer SMILES string\cite{weininger1988smiles}, from which the Polymer Structure Predictor (PSP) package\cite{sahu2022polymer} constructs a simulation box. This box undergoes a 21-step equilibration procedure\cite{abbott2013polymatic}. Subsequently, the equilibrated structures serve as the starting point for gas diffusivity and solubility calculations, accomplished through molecular dynamics and Monte Carlo simulations, respectively. Gas permeability is determined by the product of the simulated gas diffusivity and solubility.
\textbf{(c)} \textbf{Dataset overview.} Curated experimental and simulation data used for training the multi-task ML models. 
\textbf{(d)} \textbf{polyGNN}\cite{gurnani2022polymer} \textbf{architecture}. A method based on graph neural networks is initiated with a polymer SMILES string. The encoder converts the repeat unit SMILES string into a periodic graph along with fingerprints, followed by the computation of initial atomic and bond fingerprint vectors. Subsequently, the message passing unit generates the learned polymer fingerprint. Introducing a selector vector to convey data fidelity (experimental or simulation) and specific properties (permeability, diffusivity, solubility) for six gases, the approach then combines this fingerprint and selector vector before passing it to the estimator, resulting in the prediction of the desired property.
\label{fig:Overview}}
\end{figure}

A key ingredient of our MT learning approach involves simulation data that could complement measured data for gas transport, as illustrated in Fig. \ref{fig:Overview}a. To achieve this, we have designed a high-throughput molecular dynamics (MD) and Monte Carlo (MC) simulation pipeline, depicted in Figure \ref{fig:Overview}b. This pipeline generates data for gas diffusivity ($D_{sim}$) and solubility ($S_{sim}$); the subscripts explicitly indicate the source of the data. Simulated gas permeability ($P_{sim}$) is then derived from the product of $D_{sim}$ and $S_{sim}$, as prescribed by Eq. \ref{eq:pds}. Experimental data are labeled as $P_{expt}$, $D_{expt}$ and $S_{expt}$. Data for 6 different gases (CO$_{2}$, CH$_{4}$, O$_{2}$, N$_{2}$, H$_{2}$, and He) span this study. An overview of the dataset is presented in Fig. \ref{fig:Overview}c. With this fused dataset, ML models for gas transport properties are created using our newly-developed graph neural networks method – polyGNN\cite{gurnani2022polymer}, thus completing the MT learning pipeline as visualized in Figure \ref{fig:Overview}a. The input for polyGNN consists of polymer "Simplified Molecular-Input Line-Entry System" (SMILES)\cite{weininger1988smiles} strings. These SMILES strings are translated into graph representations and fingerprints, an essential ingredient for the property prediction model trained on the integrated dataset. The architecture of polyGNN, exhibited in Fig. \ref{fig:Overview}d, illustrates this process.

To test our MT learning approach, we constructed four distinct models to examine and benchmark the impact of incorporating multiple data streams. These models were designed to emulate real-world usage scenarios for the prediction model's application and to assess the improvements in prediction capabilities. To evaluate the efficacy of the MT learning, a comparison with ST learning is employed. Through these case studies, we demonstrate that MT learning surpasses conventional learning models by integrating diverse data sources and extracting meaningful correlations, particularly in data-scarce scenarios. Furthermore, the inclusion of diverse property data in this approach substantially broadens the coverage of the chemical space and effectively addresses the ML extrapolation problem. This is an ongoing process though, one that can lead to continuous improvement as more data becomes available. We then performed a head-to-head comparison of our new MT model against our previous, then state-of-the-art gas permeability predictor, deployed at Polymer Genome (\url{https://www.polymergenome.org})\cite{zhu2020polymer}, making predictions across 13 polymer classes and demonstrating the superiority of the present model. 

Finally, we highlight the power of the present development in the realm of materials discovery. Robeson-type trade-off plots are created for gas permeability, diffusivity, and solubility (by pairing each with selectivity), for over 13,000 known (i.e., previously synthesized) polymers. These trade-off plots reveal interesting candidates, as well as the true property limits across the known polymer chemical space. Most importantly the limitations of the present model (in terms of recognizing chemical spaces where the model is uncertain) are also revealed.

By integrating high-throughput simulation data with available measured data and employing data fusion techniques, one can progressively enhance the accuracy and generalizability of predictions. This philosophy and strategy holds the potential to advance polymer discovery not only for membrane technology but also for other applications.

\section{Results}

\subsection{Experimental data acquisition}

Measured gas transport properties (permeability, diffusivity, and solubility) for six different gases (CO$_{2}$, CH$_{4}$, O$_{2}$, N$_{2}$, H$_{2}$, and He) were obtained from 84 publications listed in the Polymer Handbook.\cite{brandrup1999polymer} The experimental testing temperatures ranged from 25\degree C\textendash35 \degree C, and testing pressures varied between 1\textendash30 atm. The dataset comprised a total of 820 polymers and included 3748, 709, and 550 $P_{expt}$, $D_{expt}$, and $S_{expt}$ values, respectively, amounting to a total of 5007 data points. Factors such as polymer process history and testing method were not directly included as parameters. Instead, the measured $P_{expt}$, $D_{expt}$, and $S_{expt}$ values are treated as samples from the distribution of possible values for a given polymer. As such, it is important to consider the uncertainty in the predictions, and not just the mean value of the prediction. 

\subsection{Molecular Dynamics and Monte Carlo simulations}

Gas diffusivity and solubility data were generated using classical molecular dynamics (MD) and Monte Carlo (MC) simulations, respectively. These simulations were conducted using the open-source large atomic molecular massively parallel simulator (LAMMPS) package.\cite{thompson2022lammps} The atomic potential parameters for polymers were adopted from the general AMBER force field 2 (GAFF2).\cite{wang2004development} In the simulations, the gas molecules (i.e., CO\textsubscript{2}, CH\textsubscript{4}, O\textsubscript{2}, and N\textsubscript{2}) were treated as rigid molecules, and thus were modeled with non-bonded potentials described by the TraPPE (transferable potentials for phase equilibria) models.\cite{potoff2001vapor} To perform the simulations, 27 polymer chains were inserted into the simulation box, with each chain comprising of approximately 150 atoms, and their ends were capped with a methyl group. The initial polymer configurations were generated using the Polymer Structure Predictor (PSP) package\cite{sahu2022polymer}, and a representative snapshot is shown in Fig. \ref{fig:Overview}b.

To achieve equilibrated structures, all systems underwent a 21-step relaxation procedure as recommended by Abbott et al.\cite{abbott2013polymatic} The mean-squared displacement (MSD) of the polymers was then computed, and movement beyond a few times the distance of the radius of gyration on average was assessed. This step ensured that the polymers explored various conformations and reached an equilibrium conformational state and density. Once the equilibrated structures were obtained, $D_{sim}$ and $S_{sim}$ were calculated. The simulation protocol is outlined in Fig. \ref{fig:Overview}b.

For the $D_{sim}$ calculations, a total of 27 gas molecules were randomly added to the simulation box. This specific number of molecules was chosen to be small enough to maintain the system in the dilute Fickian regime such that the gas molecules do not significantly influence each other, and yet large enough to obtain meaningful statistics. Subsequently, all systems underwent an additional equilibration of 10ns in the NPT ensemble, followed by a 100\textendash200ns production run in the NVT ensemble. The choice of a 100\textendash200ns production run duration was made to ensure the convergence of gas diffusivity and gas MSD slope across a broad spectrum of polymer types. While shorter time frames are adequate for certain instances, there are cases where the extended range of 100\textendash200ns is necessary to achieve the desired level of convergence. To illustrate this behavior, we present an analysis of simulation time versus methane diffusivity for polyethylene, polyimide, polystyrene, and polymethyl methacrylate, with the results detailed in Supplementary Figure S1. The box size in the NVT run was fixed using the average spacing and density obtained from the last 1ns of the NPT run. Nos\'e-Hoover thermostat and barostat were employed with a damping parameter of 100 time steps for each, and a time step of 1fs was used in all MD simulations. The barostat coupled the three dimensions of the box to maintain a cubic box for all systems. Simulation outputs were saved every 1000fs and block averaging from one polymer configuration was used to calculate an average $D_{sim}$ and standard deviation from the gas MSD. Block averaging allows for the reduction of random noise and more reliable statistical measures.\cite{frenkel2023understanding}

For the $S_{sim}$ calculations, a 5ns production run was performed on equilibrated structures in an NVT ensemble. During this 5ns run, a snapshot of the structure was captured every 100ps, resulting in a total of 50 snapshots. Employing an ensemble of snapshots allows for improved sampling and a standard error, which is crucial for accurate estimation of $S_{sim}$.\cite{khawaja2017molecular} Using a built-in LAMMPS function\cite{LAMMPSwidom}, 25,000 gas particles were inserted per snapshot, at random positions, following the Widom insertion method.\cite{longuet1964rigid} This method involves determining the excess chemical potential resulting from the insertion of gas molecules into the polymer, which allows for the estimation of Henry’s constant. Henry’s constant indicates how easily a particular gas dissolves in the polymer. Henry’s Law is then used to obtain gas solubility from Henry’s constant, with an assumption of a partial pressure equal to 1 atm, which is the IUPAC standard testing condition.\cite{mocak1997statistical} This derivation is detailed in the methods section. No relaxation was performed to adjust the positions of the polymer atoms or the gas particles during the insertion process. Langevin thermostat was used with a time step of 1fs for all MC simulations. 25 polymer configurations were used to calculate the $S_{sim}$, standard deviation, and the standard error from the excess chemical potential.

Fig. \ref{fig:Overview}b provides an overview of the simulation protocol used, and details of $D_{sim}$ estimation from gas MSD and $S_{sim}$ from the excess chemical potential are described in the Methods section.

\subsection{Validation of MD and MC simulations}

As an essential step of this investigation, we aimed to validate and calibrate the accuracy of the MD and MC predictions and assess the extent to which the simulations capture trends in gas transport properties. Performing classical simulations with a specific force field for polymer-gas systems across extensive chemical spaces to estimate gas diffusivity and solubility is a relatively rare endeavor. While generic force fields like GAFF2 are designed for a wide variety of materials, they often require fine-tuning of potential parameters for each unique material to attain better accuracy.

\begin{figure}[H]
    \includegraphics[width=.5\columnwidth]{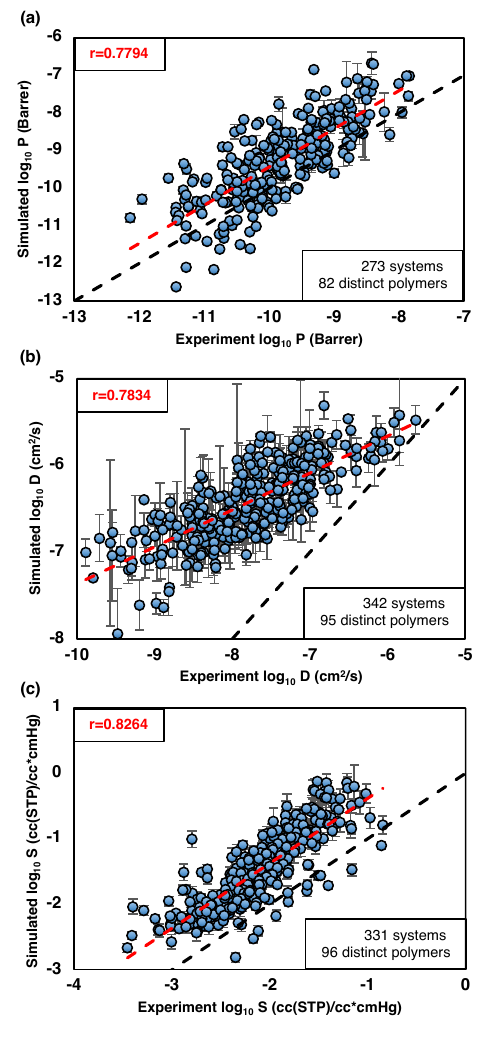}
    \caption{\textbf{(a) Gas permeability parity plot, (b) Gas diffusivity parity plot, and (c) Gas solubility parity plot.} Parity plots comparing the results from simulations against experiment data. Simulated gas permeability was derived using Eq. \ref{eq:pds}, using simulated gas diffusivity and solubility as inputs. The red lines represent trends in predicted data, while the black lines depict the parity lines of optimal fit. The error bars for all plots are represented in standard deviations. Error propagation techniques were employed to calculate the error bars for gas permeability.  While some overestimation is expected across all cases, a qualitative correlation is demonstrated.}\label{fig:parity}
\end{figure}

A total of 584 polymer-gas systems were simulated, out of which 342 systems had corresponding experimental measurements. The additional simulated systems were intended to expand the chemical coverage of the model. A comparison of $P_{sim}$, $D_{sim}$, and $S_{sim}$ against their respective experimental values, $P_{expt}$, $D_{expt}$, and $S_{expt}$, is illustrated in Fig. \ref{fig:parity}. Overall, the simulations tend to overestimate the measured values, but they effectively capture the general trends across the polymer-gas chemical space considered. The overestimation of $D_{sim}$ values, especially in low diffusivity regimes, can be attributed to the difficulty of classical force fields to accurately capture rare events and handle large chemical spaces.\cite{baba2022prediction, wang2011application} More specifically, the simulated polymer systems often exhibit lower densities compared to experimental systems, as modeled systems are approximations of the real polymeric materials and may include lower molecular weights and limited equilibration times. In our methodology, we employ a 21-step polymer equilibration relaxation procedure, which results in consistent density trends compared to experimental systems. However, a slight underestimation of density remains, as also observed by Abbott et al. in their study employing the same procedure.\cite{abbott2013polymatic} This increased free volume allows gas molecules to move more easily and quickly through the polymer system, resulting in higher diffusivity.

Similarly, the discrepancies of $S_{sim}$ relative to $S_{expt}$ may be due to the approximations inherent to the Widom insertion approach and the quality of the classical force fields across chemical spaces. Nonetheless, the favorable trends that the force fields can capture provide optimism for the usage of such simulation-derived datasets, albeit with lower fidelity, in multi-task learning frameworks. Another essential aspect of the validation is the derivation of $P_{sim}$, from the product of $D_{sim}$ and $S_{sim}$ using Eq. \ref{eq:pds}. While non-equilibrium MD can be used to simulate $P_{sim}$, it requires a more complex setup and can be computationally intensive. 

\

\subsection{Multi-task model benchmark}
 
To elucidate the effect of data fusion, we train and compare both ST and MT polyGNN models, using a subset of the experimental data collected and simulation data generated. These models were evaluated based on the predictive accuracy of $P_{expt}$, using various holdout train and test splits of 293 systems (comprised of 80 unique polymers with varying available gas data). For instance, in a 20/80 split, 20\% of the $P_{expt}$ data is set aside as testing data, while 80\% is used to train the model. To ensure representative data sampling, stratified sampling based on polymer SMILES was used when splitting the data into train and test sets. In this type of sampling, when a polymer is selected for the test set, all gas data for $P_{expt}$ associated with that polymer are withheld from the training set. This also provides insight into how well the model extrapolates to new unknown polymers. The polyGNN model training parameters used are detailed in Supplementary Table S1.

In Fig. \ref{fig:schemes}, we illustrate the two model types, ST and MT, along with the details of the train and test splits. The performance of the models was evaluated using two key metrics: the coefficient of determination (R$^2$) and the order of magnitude error (OME)\textendash units in Barrer. R$^2$ assesses how well a model predicts an outcome, while OME quantifies the prediction error in terms of orders of magnitude (taken as the logarithm of the mean absolute error). We conducted four random seed selections of the training and test sets to compute the statistics of the model performance.

\begin{figure}[H]
    \includegraphics[width=.5\columnwidth]{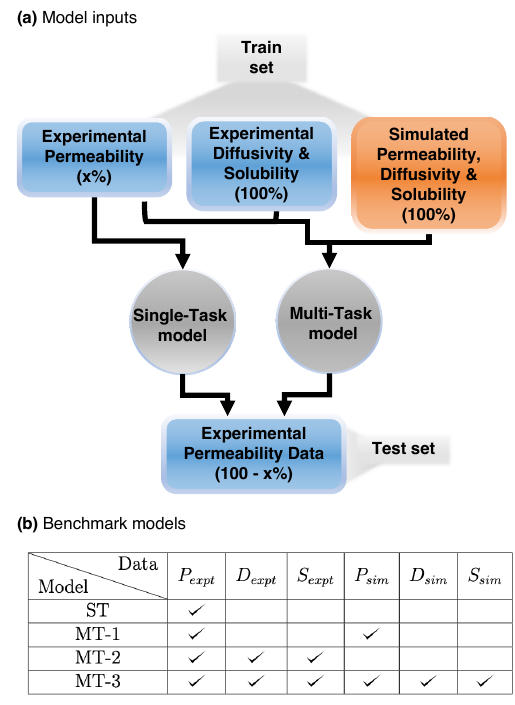}
    \caption{\textbf{(a)} \textbf{Model inputs.} This schema illustrates the train and test splits for two model variants: Regular Single Task (ST) and data-fused Multi-Task (MT) models. In the ST models, solely experimental gas permeability data is incorporated. Conversely, the MT model encompasses a possible amalgamation of experimental gas diffusivity and solubility, along with simulated gas permeability, diffusivity, and solubility data. \textbf{(b)} \textbf{Benchmark models.} Four distinct models were developed to assess the impact of MT learning. The first model (ST) exclusively incorporated experimental gas permeability data. In contrast, the subsequent MT models progressively integrated additional data. The presence of each data type in the model is indicated by the symbol "\checkmark". Here, $P$, $D$, and $S$ represent gas permeability, diffusivity, and solubility, respectively. The abbreviation $expt$ corresponds to experimental data, while $sim$ signifies simulation data.}
    \label{fig:schemes}
\end{figure}

\begin{figure}[H]
    \includegraphics[width=.5\columnwidth]{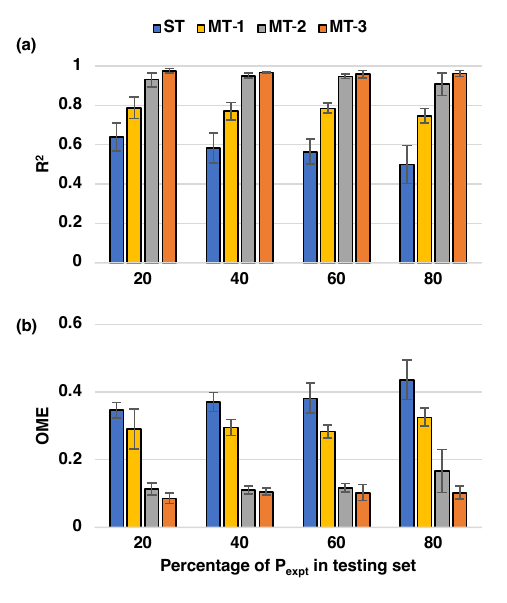}
    \caption{\textbf{Predictions of Permeability at various train test splits. (a) Coefficient of determination (R$^2$). (b) Order of magnitude error (OME).} R$^2$ evaluates the predictive performance of a model, whereas OME measures the prediction error by considering orders of magnitude, represented as the logarithm of the mean absolute error.    
    The ST and MT models are compared based on varying percentages of the unseen test set. The different test set sizes illustrate the impact of reducing training data. At 80\%, the model is trained on only 20\% of the dataset and tested on the remaining 80\%, reflecting a data-scarce region with limited chemical coverage. Comparatively, the MT models show significant improvement over the ST model, particularly at higher percentages of the unseen test set. \label{fig:results}}
\end{figure}

Our MT learning methodology comprises two primary components: the integration of simulation data and the inclusion of correlated experimental data. To establish a baseline for comparison, we employ a ST model. Shown in Fig. \ref{fig:schemes}a and represented by the "ST" row in Fig. \ref{fig:schemes}b, the ST model is exclusively trained using $P_{expt}$ data. Due to its reliance on limited data and the absence of diverse property inputs, the ST model's coverage of the chemical space is inherently constrained. As the test set percentage increases, this model is trained on progressively reduced amounts of data, leading to an anticipated decrease in predictive performance. This trend is evident in Fig. \ref{fig:results} where the R$^2$ decreases and the OME increases as the ST model is trained on diminishing data portions. In the most challenging scenario (80\% test set size), the R$^2$ dropped to less than 0.50 and the OME increased to $\approx$0.44 Barrer.

Now let's consider the first element of our MT learning approach, specifically the augmentation of $P_{expt}$ training data with $P_{sim}$, represented by the "MT-1" row in Fig. \ref{fig:schemes}. The MT-1 model is enriched with simulation data spanning the test set space. Its primary purpose is to exploit the correlations between measured and simulated data learned from the training set. This scenario mirrors situations where experimental data is unavailable, and simulation data is introduced to guide the model's predictions. Upon examining the MT-1 model, its performance noticeably surpasses that of the baseline ST model. The MT-1 model achieves an average R$^2$ and OME of $\approx$0.77 and $\approx$0.30, respectively, as shown in Fig. \ref{fig:results} (MT-1). This improvement is particularly pronounced when the test set size reaches 80\%, where the coverage of experimental data within the chemical space is most limited. This accentuates the ability of data fusion models, reinforced with simulation data, to effectively mitigate the challenges of extrapolation that conventional models (trained solely on a single experimental property) would inevitably confront. Furthermore, as another demonstration, this analysis was extended to experimental and simulation data for gas diffusivity, resulting in a similar strengthening in performance, as illustrated in Supplementary Fig. S2. This observation underlines the value of bolstering experimental data with simulation data, indicating its potential extension to other properties of interest as well.

Moving on to the second component of our MT learning methodology, we focus on augmenting the $P_{expt}$ training data with $D_{expt}$ and $S_{expt}$, represented by the "MT-2" row in Fig. \ref{fig:schemes}b. The inclusion of this supplementary data serves the purpose of empowering the model to leverage knowledge from other available pertinent properties and established physics and make predictions for the $P_{expt}$ values. In this scenario, a remarkable enhancement is observed, leading to a significant boost in predictive performance. Specifically, the average R$^2$ and OME is $\approx$0.93 and $\approx$0.12, respectively, as displayed in Fig. \ref{fig:results} (MT-2). Comparing the MT learning component in the previous passage with this second component reveals a notable difference in performance. While both approaches expand the coverage of the chemical space, MT-2 stands out due to the incorporation of high-fidelity experimental data. Unlike MT-1, where all augmented data comes from simulation, the new information in MT-2 originates from additional experimental sources, contributing to superior predictive capabilities. The MT-2 model can be likened to an ideal scenario where complementary or correlated high-fidelity data is readily available. In scenarios where such ideal conditions are not met, the MT-1 approach excels by effectively integrating simulation data to achieve a respectable level of prediction accuracy. 

In our final model, we combine the strategies embedded in both the MT-1 and MT-2 models, creating a unified model represented by row "MT-3" in Fig. \ref{fig:schemes}b. This comprehensive model encompasses all available experimental and simulation data points. The performance of the MT-3 model slightly outperforms that of the MT-2 model, exhibiting an elevated average R$^2$ of $\approx$0.96 and a comparable average OME of $\approx$0.10, as depicted in Fig. \ref{fig:results} (MT-3). Overall, this model achieves superior performance compared to the base ST model, which had an average R$^2$ and OME is $\approx$0.57 and $\approx$0.38, respectively. These results establish the efficacy of integrating simulation and correlated experimental data in successfully addressing the challenges posed by ML extrapolation.

\subsection{Production model benchmark}

\begin{figure}[H]
    \includegraphics[width=.5\columnwidth]{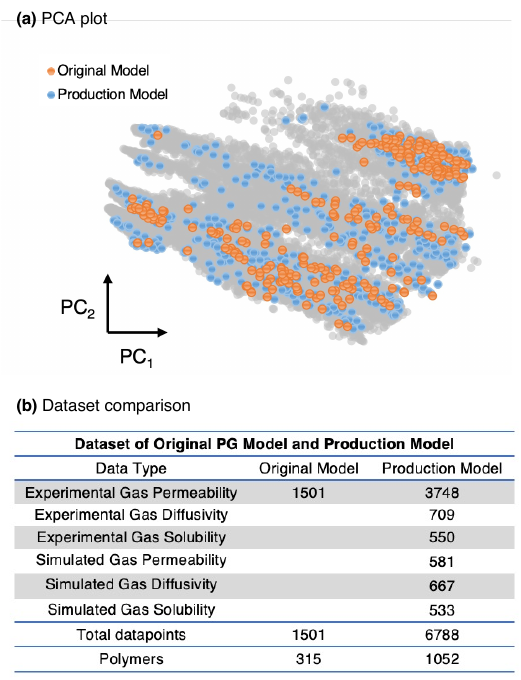}
    \caption{\textbf{(a)} \textbf{Principal Component Analysis (PCA) plot.} The PCA plot demonstrates an expanded coverage of chemical space by both the original and production models. The orange and blue dots correspond to the coverage of the original and production model, respectively, while the grey dots represent the 13,000 known polymers in our database. \textbf{(b)} \textbf{Dataset comparison.} A comparison between the original and production models reveals an incorporation of diverse data types. The production model integrates experimental and simulation data for permeability, diffusivity, and solubility properties.
    \label{fig:pca}}
\end{figure}

In the first iteration of our gas permeability prediction work, deployed at Polymer Genome (\url{https://www.polymergenome.org}), a Gaussian process regression algorithm was employed alongside a hierarchical polymer fingerprinting scheme to train a ST model.\cite{zhu2020polymer} In the present work, a transition is made to polyGNN (a recently published Graph Neural Network model that automatically generates fingerprints from SMILES strings), data augmentation, and invariant transformations to train a MT model. The models presented in the preceding section were trained using a subset of our dataset, a deliberate choice made to clearly illustrate the impact of incorporating diverse data types on prediction capabilities in a multi-task setting. Our final production model adopts the MT-3 model scheme and now incorporates all the available experimental and simulation data for gas permeability, diffusivity, and solubility. With this latest model iteration, our objective is to achieve substantial improvements over the previous version and to push the boundaries of transport predictions through polymers. The principal component analysis (PCA) plot in Fig. \ref{fig:pca}a, created using Polymer Genome fingerprints, displays the chemical space of the present study against 13,000 known polymers in our database. This plot visually demonstrates our production model's expansion to include additional chemical compositions, increasing the range of polymers for which the model can make accurate predictions. We also present a PCA plot in Supplementary Figure S3, illustrating the chemical coverage of our simulation data in comparison to experimental gas transport data and 13,000 known polymers in our database. Fig. \ref{fig:pca}b highlights the data fusion aspect of the model, showcasing the contrast between the datasets employed in the original and current models. Particularly noteworthy is the considerable enlargement of our dataset, expanding from 315 to 1052 polymers, accompanied by a significant increase in the total number of data points. With this amplified dataset, our model gains the capability to not only predict gas permeability but also include gas diffusivity and solubility. This broader scope of predictions reflects the power of our MT learning approach and its ability to leverage diverse data sources for a more comprehensive understanding of gas transport properties through polymers.

To highlight the superior performance of the present model, $P_{expt}$ predictions were made on a holdout test set of 153 systems, consisting of 31 polymers across 13 polymer classes, following a similar approach as the original model.\cite{zhu2020polymer} The summarized outcomes are presented in Table \ref{table:table2}, and the specific polymers selected for this assessment are listed in Supplementary Table S2. The overall R$^2$ has increased from 0.93 to 0.96 in the updated model compared to the original. Upon a more detailed examination of individual polymer classes, it becomes evident that the R$^2$ metric exceeds 90\% for all classes for the production model, with a particularly significant enhancement observed for polyphosphazenes, where the R$^2$ value has risen from 0.49 to 0.86. Additionally, substantial advancements have been achieved in prediction accuracy for polymers such as polynorbornenes, polypropynes, substituted polyacetylenes, and polypentynes. The diminished performance of the original model in these cases could be attributed to either limited data availability for certain polymer classes or inherent uncertainties within the experimental data. Importantly, it should be noted that the test data points for these specific polymer classes vary widely, ranging from 4 to 53 data points. This variability in data availability across diverse classes could potentially contribute to lower individual R$^2$ values for specific classes while concurrently contributing to a higher overall model R$^2$. Nonetheless, the updated model effectively overcomes these performance variations, highlighting its robustness and versatility. Further insights into the model's performance are depicted in parity plots showcasing train and test set predictions for the 31 evaluated polymers, shown in Supplementary Fig. S4 and S5.

\begin{table}[]
\centering
\begin{tabular}{lcc}
\hline
Polymer Class                                 &Original Model R$^2$  & Production Model R$^2$  \\ \hline
\rowcolor[HTML]{EFEFEF} 
Conjugated Polymers                           &\textbf{0.99} & 0.97          \\
High Temperature Polymers                     &0.94          & \textbf{0.99} \\
\rowcolor[HTML]{EFEFEF} 
Parylenes                                     &0.89          & \textbf{0.97} \\
Poly(aryl ethers) \& Poly(aryl ether ketones) &0.92          & \textbf{0.98} \\
\rowcolor[HTML]{EFEFEF} 
Polyamides \& Poly(amide-imides)              &0.96          & \textbf{0.99} \\
Polyarylates                                  &0.95          & \textbf{0.97} \\
\rowcolor[HTML]{EFEFEF} 
Polycarbonates                                &0.75          & \textbf{0.99} \\
Polyimides \& Polypyrrolones                  &\textbf{0.97} & 0.96          \\
\rowcolor[HTML]{EFEFEF} 
Polynorbornenes                               &0.51          & \textbf{0.95} \\
Polyphosphazenes                              &0.49          & \textbf{0.86} \\
\rowcolor[HTML]{EFEFEF} 
Polypropynes, polyacetylenes, polypentynes                     &0.56          & \textbf{0.94} \\
Polysulfones                                  &0.80          & \textbf{0.99} \\
\rowcolor[HTML]{EFEFEF} 
Vinyl \& Vinylidene Polymers                  &0.77          & \textbf{0.99} \\ \hline
Overall R$^2$                                    &0.93          & \textbf{0.96} \\ \hline
\end{tabular}
\caption{A comparative analysis of the predictions of the original and production models across 153 systems, encompassing 31 polymers and 13 polymer classes. The production model integrates data fusion, integrating multiple data sources and multiple properties, whereas the original model relies solely on experimentally measured gas permeability. Bold values signify the model with a superior R$^2$.    
}
\label{table:table2}
\end{table}

\section{Discussion}

\subsection{Forward-looking design}

The ideal performance of gas separation membranes is related to two intrinsic material properties: the gas permeability and the permselectivity between specific target gas pairs. Ideally, a membrane would provide high permeability and permselectivity to maximize throughput and minimize costs. In 1991, Robeson\cite{robeson1991correlation} documented a trade-off relationship between these two characteristics for polymers, often referred to as "the upper bound". This principle asserts that polymers with high permeability typically exhibit diminished selectivity, and vice versa. These upper bounds illustrate the trade-off relationship for pairs of common gases (CO$_{2}$, CH$_{4}$, O$_{2}$, N$_{2}$, H$_{2}$, and He), highlighting the best possible combination of permeability and permselectivity. This upper bound establishes a comparative benchmark for evaluating the performance metrics when designing novel membranes. As such, data-driven methods that establish a relationship between polymer structure and polymer membrane performance hold immense potential in accelerating the design of tailor-made polymers for specific separation tasks.

\begin{figure}[H]
    \includegraphics[width=.5\columnwidth]{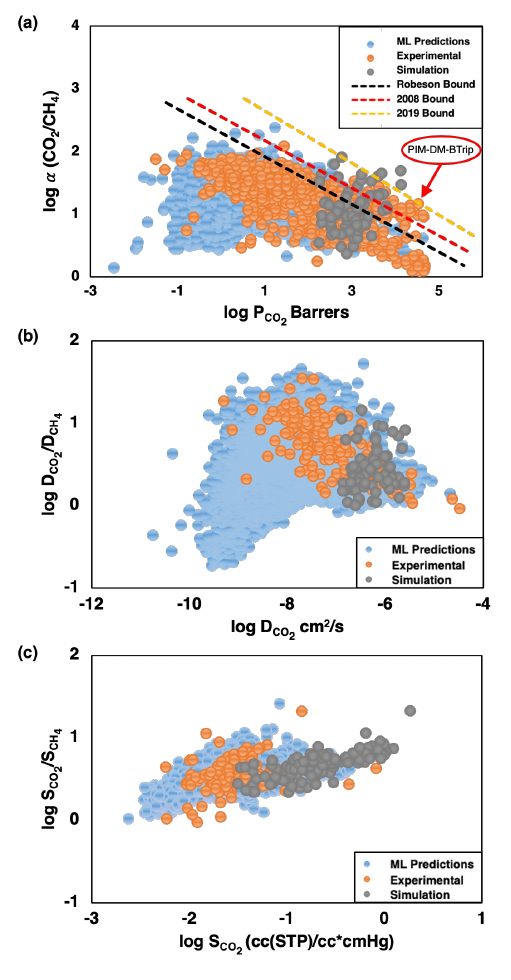}
    \caption{\textbf{Trade-off plots for \textbf{(a)} Gas permeability, \textbf{(b)} Gas diffusivity, and \textbf{(c)} Gas solubility. }
    Using our model to predict $\approx$13,000 known polymers, we compare the results to experimental data. The original Robeson upper bound (1991) and reevaluated 2008 and 2019 bounds are shown for gas permeability. There are no established bounds for gas diffusivity or solubility, but the model predictions closely align with the experimental data values. In the case of CO$_{2}$/CH$_{4}$ diffusivity selectivity, the low diffusivity regime has high uncertainty and should be taken with caution. 
        \label{fig:tradeoff}}
\end{figure}

To this end, we demonstrate our model's capability to make these assessments. We constructed permeability trade-off plots for $\approx$13,000 known polymers (i.e., previously synthesized) for the gas pairs; CO$_{2}$/CH$_{4}$, CO$_{2}$/N$_{2}$, H$_{2}$/CH$_{4}$, H$_{2}$/CO$_{2}$, O$_{2}$/N$_{2}$, and N$_{2}$/CH$_{4}$. Fig. \ref{fig:tradeoff}a shows the permeability trade-off plot for CO$_{2}$/CH$_{4}$, while the other gas pairs are shown in Supplementary Fig. S6. The ML predicted gas pair permeability and selectivity closely align with the available experimental data and the bounds, while simulation data over-predicts as expected. Both experimental and simulation data are also shown in Fig. \ref{fig:tradeoff}a. By predicting property values for the $\approx$13,000 known polymers, we can gain a clearer understanding of the overall trade-off behaviors. Robeson's upper bound, initially established in 1991, is presented alongside updated bounds introduced in 2008 and 2019.\cite{robeson2008upper,comesana2019redefining} PIM-DM-BTrip, a polymer with superior performance, is highlighted as a part of the set of polymers that helped define the 2019 bound. 

Research endeavors commonly focus on permeability trade-off plots, however as permeability can be broken down into diffusivity and solubility components, we also created CO$_{2}$/CH$_{4}$ trade-off plots for these properties, as shown in Fig. \ref{fig:tradeoff}b and Fig. \ref{fig:tradeoff}c, respectively. Diffusivity and solubility trade-off plots for CO$_{2}$/N$_{2}$, O$_{2}$/N$_{2}$, and N$_{2}$/CH$_{4}$ are illustrated in Supplementary Fig. S7 and S8. When using these models, the sensibility of predictions can be evaluated by observing common trends for the properties. For example, gas diffusivity tends to follow the relationship of $D_{O_{2}}>D_{CO_{2}}>D_{N_{2}}>D_{CH_{4}}$, a pattern primarily driven by the molecular diameter effects. However, \ref{fig:tradeoff}b, illustrates instances where the CO$_{2}$/CH$_{4}$ diffusivity selectivity falls below 1 (i.e., below 0 in the log scale). This contradicts the intuition that CO$_{2}$ diffusivity should almost always be greater than CH$_{4}$. Although there are cases where $D_{CO_{2}}$/$D_{CH_{4}}$ $<$ 1, it is a rare occurrence. A closer examination of these suspicious predictions reveals that most of them fall in the lower diffusivity regime. In this regime, prediction uncertainty, calculated using Monte Carlo dropout, tends to be inflated, revealing lower confidence for predictions. This heightened uncertainty can be directly attributed to the scarcity of data in this specific range, a challenge that is particularly pronounced in both simulations and experimental measurements. Indeed, as can be seen from \ref{fig:tradeoff}b, there are no measured or simulation data points in this property range, and hence, the ML predictions must be viewed with extreme caution and suspicion. This underscores the importance of recognizing that these models are valuable tools, but they must be used in conjunction with chemical intuition and an understanding of prediction uncertainties, especially for predictions in regions far away from the chemical space of the training set. This becomes especially critical when assessing areas with limited data or when venturing into new domains. These considerations thus mandate either experiments or simulations in such unexplored chemical spaces to better inform the ML models.
 
Trade-off plots are typically employed in designing amorphous polymers for gas separation, but when considering other applications such as packaging, the degree of crystallinity of the polymer must be considered. Gas transport behavior in semi-crystalline polymers varies due to the crystalline regions acting as impermeable barriers against gas penetration.\cite{weinkauf1990effects} Michaels et al. originally described this behavior using a two-phase model, which comprises a crystalline phase and an amorphous phase, where impedance is directly proportional to crystallinity.\cite{michaels1961solubility} Weinkauf et al. extended this model into a three-phase model that incorporates the ratio between the rigid amorphous phase fraction and mobile amorphous phase fraction (RAF/MAF).\cite{weinkauf1990effects} These models provide critical insights into the behavior of semi-crystalline polymers and offer guidelines for tailoring their gas transport properties to specific applications. The present work may be extended to address such practical situations by planning simulations of gas diffusivity and solubility through amorphous, crystalline, and amorphous-crystalline interfaces.

\subsection{Conclusion}

In this study, we introduce a novel multi-task (MT) learning approach that leverages a combination of measured and simulation data, along with correlated properties to create a state-of-the-art predictor for gas transport properties. To thoroughly evaluate the effectiveness of this approach, we performed a benchmark study in which we compared the individual impacts of each of these tasks and their collective effect when considered together. It was revealed that the addition of interrelated measured data led to a bigger benefit in enhancing the predictive capabilities of the ML model. The situation indicates that multiple correlated ground truth (i.e., measured) property data are most desirable to generate accurate property forecasts. However, in instances where rich measured data is unavailable, easily producible simulation data, when combined with measured data, demonstrates its potential by offering informed predictions. In any case, both scenarios of MT learning were able to learn underlying physical correlations and are superior to single-task (ST) models that have a less robust basis for predictions. 

These ideas have been unified to create a model representing a major advancement in predicting gas transport properties through polymers. This model in a comparative analysis with the prior work displayed concrete improvements, across 13 different polymer classes. Using our new ML model we also generate selectivity trade-off plots for gas permeability, diffusivity, and solubility for $\approx$13,000 known polymers (i.e., previously synthesized). These plots provide insights into the strengths and limitations of the modes, but more importantly, the need for data across diverse chemical spaces, e.g., via simulations if measured data proves to be laborious to generate. The prospect of continual expansion of the accessible polymer universe will push the frontiers of what is achievable in terms of properties and performance. 

\section{Methods}

\subsection{Gas diffusivity calculation in MD simulations}

From our MD simulations, the diffusivity ($D_{sim}$) of gas molecules was obtained by:
\begin{equation}
D_{sim}=\frac{1}{6N_{gas}}\lim_{t\to\infty}\frac{d}{dt}\sum_{i=1}^{N_{gas}}\bigl\langle\Delta r_{i}(t)^{2}\bigr\rangle\label{eq:diffusion_Einstein}
\end{equation}
where $N_{gas}$ is the number of gas molecules in the simulation cell, $t$ is the simulation time, $r_{i}(t)$ is the position of the gas molecule $i$ at time $t$, $\Delta r_{i}(t) = r_{i}(t) - r_{i}(0)$, is the displacement of gas $i$ between time $0$ and time $t$, and $\bigl\langle\Delta r_{i}(t)^{2}\bigr\rangle$ is the mean square displacement (MSD) of gas molecule $i$ at time $t$. The gas MSD was block averaged over 2-5 non-overlapping trajectories depending on the system dynamics (depending on whether breaking the trajectory into shorter blocks still allows its MSD curve to reach the Fickian regime). This block averaging was also used to calculate the standard deviation. The diffusivity was obtained using the slope from a least-squares linear fit of the final decade of the MSD data. The log-log slopes of all the systems' MSD curves are in the range of 0.95-1.05, which is a common range for classical MD simulation diffusivity studies.\cite{shen2018diffusion, shen2020effects, shen2020ion}

\subsection{Gas solubility calculation in MC simulations}
 
The Widom Insertion method was used to calculate the Henry's constant ($k$) of gas molecules within a Monte Carlo simulation. In this method, we insert N gas molecules into the simulation box (one at a time, at various random locations), and the excess chemical potential ($\mu_{ex}$) of the gas in the membrane is obtained. The estimation uses an ensemble average of the $N$ separate, random, insertions, the $i$th of which will change the internal energy of the system by $\Delta E_{i}$. 
\begin{equation}
\mu_{ex}={-k_{b}T}\ln\langle\exp\left(\frac{-\Delta E_{i}}{k_{b}T}\right)\rangle_{N}\label{eq:uex}
\end{equation}
where $k_{b}$ is Boltzmann constant, $T$ is temperature and $N$ = 25,000 insertions. This estimation is defined for the dilute limit, where there are no gas-gas interactions. When an insertion overlaps with the polymer in the system, the $\mu_{ex}$ results in minimal contribution towards gas solubility, and thus these insertions with energies greater than 5$k_{B}T$ are discarded. With $\mu_{ex}$, we can obtain Henry's constant, $k$, of gas molecules by:
\begin{equation}
k =\exp\left(\frac{\mu_{ex}}{k_{b}T}\right)_{N}\label{eq:solubility}
\end{equation}

With $k$, we then use Henry's law which states that the solubility ($S$) of a gas is directly proportional to the partial pressure of the gas ($P_{gas}$), which takes the form:
\begin{equation}
S =kP_{gas}\label{eq:hc2}
\end{equation}
We assume a standard testing condition of partial pressure equal to 1 atm, and thus:
\begin{equation}
S=k\label{eq:ssim}
\end{equation}
In this work, for each polymer configuration, we take $m$=50 dynamic snapshots from a 5ns production run on the equilibrated structure in an NVT ensemble. An average solubility, $S_{i}$ of the configuration is obtained by: 

\begin{equation}
S_{i}=  \frac{1}{m} \sum_{i=1}^{m}  S_{ij}\label{eq:ssim}
\end{equation}

For proper convergence and as a measure of standard error, we use $n$ = 25 polymer configurations and $S_{sim}$ is obtained by:
\begin{equation}
S_{sim}=\frac{1}{n} \sum_{i=1}^{n}  S_{i}\label{eq:ssim}
\end{equation}
The standard deviation, $\sigma$, and standard error, SE, are then calculated as: 
\begin{equation}
\sigma^2 =\frac{1}{n} \sum_{i=1}^{n} \left(S_{i}-S_{sim}\right)^2\label{eq:stdev}
\end{equation}
\begin{equation}
SE =\frac{\sigma}{\sqrt{n}} \label{eq:se}
\end{equation}
Final gas solubility values were screened for a SE of less than 5$\%$.

\subsection{polyGNN}

The predictive model we used was polyGNN, a multitask graph neural network method that has shown promising results when dealing with large-scale multi-property datasets.\cite{gurnani2022polymer} Briefly, polyGNN contains three modules: the Encoder, Message Passing Block, and the Estimator. The inputs to polyGNN are a polymer repeat unit and a property of interest (or, equivalently, the property's associated selector vector). The two outputs of a polyGNN model are the repeat unit's fingerprint and the value of the property of interest.
In the Encoder, the repeat unit is first converted to a periodic graph, with each atom as a node and each bond as an edge. Then, each node and edge in the graph are given an initial fingerprint. After the graph elements have been assigned their initial features, the graph is passed to the Message Passing Block. Messages between neighboring atoms are iteratively passed along chemical bonds. After each iteration, every node fingerprint is updated using the messages, while each bond fingerprint remains the same. The message passed from atom $j$ to atom $i$ at time step $k$ is calculated according to Equation \ref{eq:mij}.

\begin{equation}
m_{i,j}^{(k)}=\phi^{(k)}\left(x_{i}^{(k)},x_{j}^{(k)},e_{i,j}\right)\label{eq:mij}
\end{equation}
where each $\phi^{(k)}$ is a parameterized function, $x_{i}^{(k)}$ and $x_{j}^{(k)}$ are the encodings of neighboring ij atoms after time step $k$, and $e_{i,j}$ is the fingerprint of the bond that joins atoms $i$, $j$. $m_{i,j}^{(k)} = 0$ if $i$, $j$ do not share a chemical bond. After initialization, each node receives messages from all of its neighbors. These messages are aggregated by some permutation-invariant function $f$ (e.g., sum, mean, max). We use the sum in this work. The aggregated message, along with the current node encoding, is used to update the node encoding. The node update process is defined in Equation \ref{eq:xi}.

\begin{equation}
x_{i}^{(k)}=\chi^{(k)}\left(\chi_{i}^{(k-1)},f\left(\left\{ m_{i,j}\forall j\in [1,N_{p}] \right\} \right)\right)+x_{i}^{(k-2)}\label{eq:xi}
\end{equation}
where each $\chi^{(k)}$ is a parameterized function, $p$ is a polymer, $[1,Np]$ is the set of integers between 1 and $N_p$, $N_p$ is the number of atoms in the repeat unit of $p$, and $x^{(k)} = 0, \forall k < 0$. Messages are passed for $\tau$ time steps, where $\tau$ is also the capacity in this work. The fingerprint of the entire polymer, $x_p$, is calculated by the graph aggregation function $A_g$, as shown in Equation \ref{eq:xp}.

\begin{equation}
x_{p}=A_{g}\left(x_{i}^{(\tau)},x_{i}^{(0)}\right)=\frac{1}{N_{p}} \sum_{i=1}^{N_{p}}x_{i}^{(\tau)}+x_{i}^{(0)}\label{eq:xp}
\end{equation}

Finally, $x_p$ and the selector $s$ can be passed to the Estimator. Here, these inputs are mapped to a polymer property prediction, $y_p$, via a parameterized function $\psi$, which represents the multilayer perceptron (MLP) depth.

\begin{equation}
y_{p}=\psi\left(x_{p},s\right)\label{eq:yp}
\end{equation}

$\psi$ specifies the number of hidden layers between the input and output layers, with this depth parameterized to range from 2 to 14 layers. During training, the parameters of all $\phi (k)$, $\chi (k)$, $\psi$ are learned simultaneously. As shown in Equation \ref{eq:xi}, our update step leverages skip connections, which have been shown to improve the optimization of shallow layers in deep neural networks.

All neural network architectures used dropout layers, fully connected layers, and Leaky ReLU activations (with a negative slope equal to 0.01). MC dropout was implemented by performing 10 forward passes through the network, each time applying dropout to different subsets of nodes. All architectures were created using PyTorch and PyTorch Geometric. The weights of all models were optimized using the Adam optimizer and the mean squared error loss function. 

\subsection{Training procedure}

The training procedure used is similar to that in the polyGNN work, where the models are ensemble models, composed of several submodels.\cite{gurnani2022polymer} The output of the ensemble is computed by the average of each submodel's output. The data used for training was grouped based on gas transport type ($P$, $D$, \& $S$), gas type, and data source (experiment or simulation). Once grouped, each data subset was then min-max scaled between 0 and 1. The polyGNN model training parameters used are detailed below and also compiled in Supplementary Table S1.

Next, the entire data set was stratified and split into training and test sets (percentages of test sets were 20\%, 40\%, 60\%, or 80\%) based on polymer SMILES strings three times. Using the NNDebugger package,\cite{gurnani2022debugging} the optimal capacity was found by attempting to overfit (R$^2$$>0.97$) the entire training data set. If the data was not overfit, then the capacity corresponding to the highest R$^2$ value was used. The capacity range considered was between two and fourteen. The training data set was then divided into an 80\% hyperparameter (HP) training set and a 20\% HP validation set. The remaining HPs (batch size, learning rate, dropout percentage) were optimized using the package scikit-optimize.\cite{head2018scikit} The set of HPs corresponding to the lowest RMSE on the HP validation set was considered optimal.

Finally, the training data set was split into five folds using cross-validation (CV), producing one CV train data set and one CV validation data set per fold. For each fold, the model's HPs were fixed as the optimal HPs and the model's learnable parameters were fit to the CV train data set for 1000 epochs. At the end of 1000 epochs, the model parameters corresponding to the epoch with the lowest RMSE in the CV validation data set were chosen. After all five models were trained on their respective CV splits, the models were placed in an ensemble. The ensemble was used to make predictions of the test set, that were completely unseen by the ensemble during HP optimization or model training with CV.

\section{Data availability}

The experimental sources of data used are reported in the paper. All data, experimental and simulation, are available free of charge at \url{https://github.com/Ramprasad-Group/polyVERSE/tree/main/Other/Gas_permeability_solubility_diffusivity}.

\section{Code availability}

The Polymer Structure Predictor (PSP) package to create simulation polymer structures is available free of charge at \url{https://github.com/Ramprasad-Group/PSP}

The code used to perform molecular dynamics (MD) and Monte Carlo (MC) simulations is available free of charge at \url{https://github.com/Ramprasad-Group/polyVERSE/tree/main/Other/Gas_permeability_solubility_diffusivity}. 

The code used to train our polyGNN models is available at \url{https://github.com/Ramprasad-Group/polygnn} for academic use.

\section{Acknowledgments}
This work is financially supported by Toyota Research Institute through the Accelerated Materials Design and Discovery program and the Office of Naval Research through a multidisciplinary university research initiative (MURI) grant N00014-20-1−2586. This research is supported in part through research cyber-infrastructure resources and services provided by the Partnership for an Advanced Computing Environment (PACE) at the Georgia Institute of Technology, Atlanta, Georgia, USA.\cite{PACE} The authors thank XSEDE/ACCESS for computational support through Grant No. TG-DMR080058N.

\newpage
\bibliography{refs}

\providecommand{\latin}[1]{#1}
\makeatletter
\providecommand{\doi}
  {\begingroup\let\do\@makeother\dospecials
  \catcode`\{=1 \catcode`\}=2 \doi@aux}
\providecommand{\doi@aux}[1]{\endgroup\texttt{#1}}
\makeatother
\providecommand*\mcitethebibliography{\thebibliography}
\csname @ifundefined\endcsname{endmcitethebibliography}  {\let\endmcitethebibliography\endthebibliography}{}
\begin{mcitethebibliography}{56}
\providecommand*\natexlab[1]{#1}
\providecommand*\mciteSetBstSublistMode[1]{}
\providecommand*\mciteSetBstMaxWidthForm[2]{}
\providecommand*\mciteBstWouldAddEndPuncttrue
  {\def\EndOfBibitem{\unskip.}}
\providecommand*\mciteBstWouldAddEndPunctfalse
  {\let\EndOfBibitem\relax}
\providecommand*\mciteSetBstMidEndSepPunct[3]{}
\providecommand*\mciteSetBstSublistLabelBeginEnd[3]{}
\providecommand*\EndOfBibitem{}
\mciteSetBstSublistMode{f}
\mciteSetBstMaxWidthForm{subitem}{(\alph{mcitesubitemcount})}
\mciteSetBstSublistLabelBeginEnd
  {\mcitemaxwidthsubitemform\space}
  {\relax}
  {\relax}

\bibitem[Ferreira \latin{et~al.}(2016)Ferreira, Alves, and Coelhoso]{Ferreira2016Polysaccharide}
Ferreira,~A.; Alves,~V.; Coelhoso,~I. Polysaccharide-based membranes in food packaging applications. \emph{Membranes} \textbf{2016}, \emph{6}, 22\relax
\mciteBstWouldAddEndPuncttrue
\mciteSetBstMidEndSepPunct{\mcitedefaultmidpunct}
{\mcitedefaultendpunct}{\mcitedefaultseppunct}\relax
\EndOfBibitem
\bibitem[Baker(2001)]{Baker_2001Membrane}
Baker,~R.~W. Membrane technology. \emph{Encyclopedia of Polymer Science and Technology} \textbf{2001}, \relax
\mciteBstWouldAddEndPunctfalse
\mciteSetBstMidEndSepPunct{\mcitedefaultmidpunct}
{}{\mcitedefaultseppunct}\relax
\EndOfBibitem
\bibitem[Wijmans and Baker(1995)Wijmans, and Baker]{Wijmans_Baker_1995}
Wijmans,~J.; Baker,~R. The solution-diffusion model: A Review. \emph{Journal of Membrane Science} \textbf{1995}, \emph{107}, 1–21\relax
\mciteBstWouldAddEndPuncttrue
\mciteSetBstMidEndSepPunct{\mcitedefaultmidpunct}
{\mcitedefaultendpunct}{\mcitedefaultseppunct}\relax
\EndOfBibitem
\bibitem[Tran \latin{et~al.}(2023)Tran, Shen, Shukla, Kwon, and Ramprasad]{tran2023informatics2}
Tran,~H.; Shen,~K.-H.; Shukla,~S.; Kwon,~H.-K.; Ramprasad,~R. Informatics-Driven Selection of Polymers for Fuel-Cell Applications. \emph{The Journal of Physical Chemistry C} \textbf{2023}, \emph{127}, 977--986\relax
\mciteBstWouldAddEndPuncttrue
\mciteSetBstMidEndSepPunct{\mcitedefaultmidpunct}
{\mcitedefaultendpunct}{\mcitedefaultseppunct}\relax
\EndOfBibitem
\bibitem[Barnett \latin{et~al.}(2020)Barnett, Bilchak, Wang, Benicewicz, Murdock, Bereau, and Kumar]{barnett2020designing}
Barnett,~J.~W.; Bilchak,~C.~R.; Wang,~Y.; Benicewicz,~B.~C.; Murdock,~L.~A.; Bereau,~T.; Kumar,~S.~K. Designing exceptional gas-separation polymer membranes using machine learning. \emph{Science Advances} \textbf{2020}, \emph{6}, eaaz4301\relax
\mciteBstWouldAddEndPuncttrue
\mciteSetBstMidEndSepPunct{\mcitedefaultmidpunct}
{\mcitedefaultendpunct}{\mcitedefaultseppunct}\relax
\EndOfBibitem
\bibitem[Moore \latin{et~al.}(2004)Moore, Damle, Williams, and Koros]{moore2004characterization}
Moore,~T.~T.; Damle,~S.; Williams,~P.~J.; Koros,~W.~J. Characterization of low permeability gas separation membranes and barrier materials; design and operation considerations. \emph{Journal of Membrane Science} \textbf{2004}, \emph{245}, 227--231\relax
\mciteBstWouldAddEndPuncttrue
\mciteSetBstMidEndSepPunct{\mcitedefaultmidpunct}
{\mcitedefaultendpunct}{\mcitedefaultseppunct}\relax
\EndOfBibitem
\bibitem[M{\"u}ller-Plathe(1994)]{Müller1994}
M{\"u}ller-Plathe,~F. Permeation of polymers—a computational approach. \emph{Acta Polymerica} \textbf{1994}, \emph{45}, 259--293\relax
\mciteBstWouldAddEndPuncttrue
\mciteSetBstMidEndSepPunct{\mcitedefaultmidpunct}
{\mcitedefaultendpunct}{\mcitedefaultseppunct}\relax
\EndOfBibitem
\bibitem[Audus and de~Pablo(2017)Audus, and de~Pablo]{audus2017polymer}
Audus,~D.~J.; de~Pablo,~J.~J. Polymer informatics: Opportunities and challenges. \emph{ACS Macro Letters} \textbf{2017}, \emph{6}, 1078--1082\relax
\mciteBstWouldAddEndPuncttrue
\mciteSetBstMidEndSepPunct{\mcitedefaultmidpunct}
{\mcitedefaultendpunct}{\mcitedefaultseppunct}\relax
\EndOfBibitem
\bibitem[Batra \latin{et~al.}(2021)Batra, Song, and Ramprasad]{batra2021emerging}
Batra,~R.; Song,~L.; Ramprasad,~R. Emerging materials intelligence ecosystems propelled by machine learning. \emph{Nature Reviews Materials} \textbf{2021}, \emph{6}, 655--678\relax
\mciteBstWouldAddEndPuncttrue
\mciteSetBstMidEndSepPunct{\mcitedefaultmidpunct}
{\mcitedefaultendpunct}{\mcitedefaultseppunct}\relax
\EndOfBibitem
\bibitem[Chen \latin{et~al.}(2021)Chen, Pilania, Batra, Huan, Kim, Kuenneth, and Ramprasad]{chen2021polymer}
Chen,~L.; Pilania,~G.; Batra,~R.; Huan,~T.~D.; Kim,~C.; Kuenneth,~C.; Ramprasad,~R. Polymer informatics: Current status and critical next steps. \emph{Materials Science and Engineering: R: Reports} \textbf{2021}, \emph{144}, 100595\relax
\mciteBstWouldAddEndPuncttrue
\mciteSetBstMidEndSepPunct{\mcitedefaultmidpunct}
{\mcitedefaultendpunct}{\mcitedefaultseppunct}\relax
\EndOfBibitem
\bibitem[Zhu \latin{et~al.}(2020)Zhu, Kim, Chandrasekarn, Everett, Ramprasad, and Lively]{zhu2020polymer}
Zhu,~G.; Kim,~C.; Chandrasekarn,~A.; Everett,~J.~D.; Ramprasad,~R.; Lively,~R.~P. Polymer genome--based prediction of gas permeabilities in polymers. \emph{Journal of Polymer Engineering} \textbf{2020}, \emph{40}, 451--457\relax
\mciteBstWouldAddEndPuncttrue
\mciteSetBstMidEndSepPunct{\mcitedefaultmidpunct}
{\mcitedefaultendpunct}{\mcitedefaultseppunct}\relax
\EndOfBibitem
\bibitem[Wu \latin{et~al.}(2022)Wu, Deshmukh, Chen, Ramprasad, Sotzing, and Cao]{wu2022rational}
Wu,~C.; Deshmukh,~A.~A.; Chen,~L.; Ramprasad,~R.; Sotzing,~G.~A.; Cao,~Y. Rational design of all-organic flexible high-temperature polymer dielectrics. \emph{Matter} \textbf{2022}, \emph{5}, 2615--2623\relax
\mciteBstWouldAddEndPuncttrue
\mciteSetBstMidEndSepPunct{\mcitedefaultmidpunct}
{\mcitedefaultendpunct}{\mcitedefaultseppunct}\relax
\EndOfBibitem
\bibitem[Chen \latin{et~al.}(2020)Chen, Kim, Batra, Lightstone, Wu, Li, Deshmukh, Wang, Tran, Vashishta, \latin{et~al.} others]{chen2020frequency}
Chen,~L.; Kim,~C.; Batra,~R.; Lightstone,~J.~P.; Wu,~C.; Li,~Z.; Deshmukh,~A.~A.; Wang,~Y.; Tran,~H.~D.; Vashishta,~P.; others Frequency-dependent dielectric constant prediction of polymers using machine learning. \emph{npj Computational Materials} \textbf{2020}, \emph{6}, 61\relax
\mciteBstWouldAddEndPuncttrue
\mciteSetBstMidEndSepPunct{\mcitedefaultmidpunct}
{\mcitedefaultendpunct}{\mcitedefaultseppunct}\relax
\EndOfBibitem
\bibitem[Wessling \latin{et~al.}(1994)Wessling, Mulder, Bos, Van Der~Linden, Bos, and Van Der~Linden]{wessling1994modelling}
Wessling,~M.; Mulder,~M.; Bos,~A.; Van Der~Linden,~M.; Bos,~M.; Van Der~Linden,~W. Modelling the permeability of polymers: a neural network approach. \emph{Journal of membrane science} \textbf{1994}, \emph{86}, 193--198\relax
\mciteBstWouldAddEndPuncttrue
\mciteSetBstMidEndSepPunct{\mcitedefaultmidpunct}
{\mcitedefaultendpunct}{\mcitedefaultseppunct}\relax
\EndOfBibitem
\bibitem[Yuan \latin{et~al.}(2021)Yuan, Longo, Thornton, McKeown, Comesana-Gandara, Jansen, and Jelfs]{yuan2021imputation}
Yuan,~Q.; Longo,~M.; Thornton,~A.~W.; McKeown,~N.~B.; Comesana-Gandara,~B.; Jansen,~J.~C.; Jelfs,~K.~E. Imputation of missing gas permeability data for polymer membranes using machine learning. \emph{Journal of Membrane Science} \textbf{2021}, \emph{627}, 119207\relax
\mciteBstWouldAddEndPuncttrue
\mciteSetBstMidEndSepPunct{\mcitedefaultmidpunct}
{\mcitedefaultendpunct}{\mcitedefaultseppunct}\relax
\EndOfBibitem
\bibitem[Ricci and De~Angelis(2023)Ricci, and De~Angelis]{ricci2023perspective}
Ricci,~E.; De~Angelis,~M.~G. A perspective on data-driven screening and discovery of polymer membranes for gas separation, from the molecular structure to the industrial performance. \emph{Reviews in Chemical Engineering} \textbf{2023}, \relax
\mciteBstWouldAddEndPunctfalse
\mciteSetBstMidEndSepPunct{\mcitedefaultmidpunct}
{}{\mcitedefaultseppunct}\relax
\EndOfBibitem
\bibitem[Wang \latin{et~al.}(2006)Wang, Shao, Wang, and Wu]{wang2006radial}
Wang,~L.; Shao,~C.; Wang,~H.; Wu,~H. Radial basis function neural networks-based modeling of the membrane separation process: hydrogen recovery from refinery gases. \emph{Journal of Natural Gas Chemistry} \textbf{2006}, \emph{15}, 230--234\relax
\mciteBstWouldAddEndPuncttrue
\mciteSetBstMidEndSepPunct{\mcitedefaultmidpunct}
{\mcitedefaultendpunct}{\mcitedefaultseppunct}\relax
\EndOfBibitem
\bibitem[Rogers and Hahn(2010)Rogers, and Hahn]{rogers2010extended}
Rogers,~D.; Hahn,~M. Extended-connectivity fingerprints. \emph{Journal of chemical information and modeling} \textbf{2010}, \emph{50}, 742--754\relax
\mciteBstWouldAddEndPuncttrue
\mciteSetBstMidEndSepPunct{\mcitedefaultmidpunct}
{\mcitedefaultendpunct}{\mcitedefaultseppunct}\relax
\EndOfBibitem
\bibitem[Landrum(2013)]{landrum2013rdkit}
Landrum,~G. Rdkit documentation. \emph{Release} \textbf{2013}, \emph{1}, 4\relax
\mciteBstWouldAddEndPuncttrue
\mciteSetBstMidEndSepPunct{\mcitedefaultmidpunct}
{\mcitedefaultendpunct}{\mcitedefaultseppunct}\relax
\EndOfBibitem
\bibitem[Huan \latin{et~al.}(2015)Huan, Mannodi-Kanakkithodi, and Ramprasad]{huan2015accelerated}
Huan,~T.~D.; Mannodi-Kanakkithodi,~A.; Ramprasad,~R. Accelerated materials property predictions and design using motif-based fingerprints. \emph{Physical Review B} \textbf{2015}, \emph{92}, 014106\relax
\mciteBstWouldAddEndPuncttrue
\mciteSetBstMidEndSepPunct{\mcitedefaultmidpunct}
{\mcitedefaultendpunct}{\mcitedefaultseppunct}\relax
\EndOfBibitem
\bibitem[Le \latin{et~al.}(2012)Le, Epa, Burden, and Winkler]{le2012quantitative}
Le,~T.; Epa,~V.~C.; Burden,~F.~R.; Winkler,~D.~A. Quantitative structure--property relationship modeling of diverse materials properties. \emph{Chemical Reviews} \textbf{2012}, \emph{112}, 2889--2919\relax
\mciteBstWouldAddEndPuncttrue
\mciteSetBstMidEndSepPunct{\mcitedefaultmidpunct}
{\mcitedefaultendpunct}{\mcitedefaultseppunct}\relax
\EndOfBibitem
\bibitem[Gurnani \latin{et~al.}(2022)Gurnani, Kuenneth, Toland, and Ramprasad]{gurnani2022polymer}
Gurnani,~R.; Kuenneth,~C.; Toland,~A.; Ramprasad,~R. Polymer informatics at-scale with multitask graph neural networks. \emph{arXiv preprint arXiv:2209.13557} \textbf{2022}, \relax
\mciteBstWouldAddEndPunctfalse
\mciteSetBstMidEndSepPunct{\mcitedefaultmidpunct}
{}{\mcitedefaultseppunct}\relax
\EndOfBibitem
\bibitem[Kuenneth and Ramprasad(2023)Kuenneth, and Ramprasad]{kuenneth2023polybert}
Kuenneth,~C.; Ramprasad,~R. polyBERT: a chemical language model to enable fully machine-driven ultrafast polymer informatics. \emph{Nature Communications} \textbf{2023}, \emph{14}, 4099\relax
\mciteBstWouldAddEndPuncttrue
\mciteSetBstMidEndSepPunct{\mcitedefaultmidpunct}
{\mcitedefaultendpunct}{\mcitedefaultseppunct}\relax
\EndOfBibitem
\bibitem[Ramprasad \latin{et~al.}(2017)Ramprasad, Batra, Pilania, Mannodi-Kanakkithodi, and Kim]{ramprasad2017machine}
Ramprasad,~R.; Batra,~R.; Pilania,~G.; Mannodi-Kanakkithodi,~A.; Kim,~C. Machine learning in materials informatics: recent applications and prospects. \emph{npj Computational Materials} \textbf{2017}, \emph{3}, 54\relax
\mciteBstWouldAddEndPuncttrue
\mciteSetBstMidEndSepPunct{\mcitedefaultmidpunct}
{\mcitedefaultendpunct}{\mcitedefaultseppunct}\relax
\EndOfBibitem
\bibitem[Hutchinson \latin{et~al.}(2017)Hutchinson, Antono, Gibbons, Paradiso, Ling, and Meredig]{hutchinson2017overcoming}
Hutchinson,~M.~L.; Antono,~E.; Gibbons,~B.~M.; Paradiso,~S.; Ling,~J.; Meredig,~B. Overcoming data scarcity with transfer learning. \emph{arXiv preprint arXiv:1711.05099} \textbf{2017}, \relax
\mciteBstWouldAddEndPunctfalse
\mciteSetBstMidEndSepPunct{\mcitedefaultmidpunct}
{}{\mcitedefaultseppunct}\relax
\EndOfBibitem
\bibitem[Caruana(1998)]{caruana1998multitask}
Caruana,~R. \emph{Learning to Learn}; Springer, 1998\relax
\mciteBstWouldAddEndPuncttrue
\mciteSetBstMidEndSepPunct{\mcitedefaultmidpunct}
{\mcitedefaultendpunct}{\mcitedefaultseppunct}\relax
\EndOfBibitem
\bibitem[Patra \latin{et~al.}(2020)Patra, Batra, Chandrasekaran, Kim, Huan, and Ramprasad]{patra2020multi}
Patra,~A.; Batra,~R.; Chandrasekaran,~A.; Kim,~C.; Huan,~T.~D.; Ramprasad,~R. A multi-fidelity information-fusion approach to machine learn and predict polymer bandgap. \emph{Computational Materials Science} \textbf{2020}, \emph{172}, 109286\relax
\mciteBstWouldAddEndPuncttrue
\mciteSetBstMidEndSepPunct{\mcitedefaultmidpunct}
{\mcitedefaultendpunct}{\mcitedefaultseppunct}\relax
\EndOfBibitem
\bibitem[Kuenneth \latin{et~al.}(2021)Kuenneth, Rajan, Tran, Chen, Kim, and Ramprasad]{kuenneth2021polymer}
Kuenneth,~C.; Rajan,~A.~C.; Tran,~H.; Chen,~L.; Kim,~C.; Ramprasad,~R. Polymer informatics with multi-task learning. \emph{Patterns} \textbf{2021}, \emph{2}, 100238\relax
\mciteBstWouldAddEndPuncttrue
\mciteSetBstMidEndSepPunct{\mcitedefaultmidpunct}
{\mcitedefaultendpunct}{\mcitedefaultseppunct}\relax
\EndOfBibitem
\bibitem[Yang \latin{et~al.}(2022)Yang, Tao, He, McCutcheon, and Li]{yang2022machine}
Yang,~J.; Tao,~L.; He,~J.; McCutcheon,~J.~R.; Li,~Y. Machine learning enables interpretable discovery of innovative polymers for gas separation membranes. \emph{Science Advances} \textbf{2022}, \emph{8}, eabn9545\relax
\mciteBstWouldAddEndPuncttrue
\mciteSetBstMidEndSepPunct{\mcitedefaultmidpunct}
{\mcitedefaultendpunct}{\mcitedefaultseppunct}\relax
\EndOfBibitem
\bibitem[Venkatram \latin{et~al.}(2020)Venkatram, Batra, Chen, Kim, Shelton, and Ramprasad]{venkatram2020predicting}
Venkatram,~S.; Batra,~R.; Chen,~L.; Kim,~C.; Shelton,~M.; Ramprasad,~R. Predicting crystallization tendency of polymers using multifidelity information fusion and machine learning. \emph{The Journal of Physical Chemistry B} \textbf{2020}, \emph{124}, 6046--6054\relax
\mciteBstWouldAddEndPuncttrue
\mciteSetBstMidEndSepPunct{\mcitedefaultmidpunct}
{\mcitedefaultendpunct}{\mcitedefaultseppunct}\relax
\EndOfBibitem
\bibitem[Weininger(1988)]{weininger1988smiles}
Weininger,~D. SMILES, a chemical language and information system. 1. Introduction to methodology and encoding rules. \emph{Journal of Chemical Information and Computer Sciences} \textbf{1988}, \emph{28}, 31--36\relax
\mciteBstWouldAddEndPuncttrue
\mciteSetBstMidEndSepPunct{\mcitedefaultmidpunct}
{\mcitedefaultendpunct}{\mcitedefaultseppunct}\relax
\EndOfBibitem
\bibitem[Sahu \latin{et~al.}(2022)Sahu, Shen, Montoya, Tran, and Ramprasad]{sahu2022polymer}
Sahu,~H.; Shen,~K.-H.; Montoya,~J.~H.; Tran,~H.; Ramprasad,~R. Polymer structure predictor (psp): a python toolkit for predicting atomic-level structural models for a range of polymer geometries. \emph{Journal of Chemical Theory and Computation} \textbf{2022}, \emph{18}, 2737--2748\relax
\mciteBstWouldAddEndPuncttrue
\mciteSetBstMidEndSepPunct{\mcitedefaultmidpunct}
{\mcitedefaultendpunct}{\mcitedefaultseppunct}\relax
\EndOfBibitem
\bibitem[Abbott \latin{et~al.}(2013)Abbott, Hart, and Colina]{abbott2013polymatic}
Abbott,~L.~J.; Hart,~K.~E.; Colina,~C.~M. Polymatic: a generalized simulated polymerization algorithm for amorphous polymers. \emph{Theoretical Chemistry Accounts} \textbf{2013}, \emph{132}, 1--19\relax
\mciteBstWouldAddEndPuncttrue
\mciteSetBstMidEndSepPunct{\mcitedefaultmidpunct}
{\mcitedefaultendpunct}{\mcitedefaultseppunct}\relax
\EndOfBibitem
\bibitem[Brandrup \latin{et~al.}(1999)Brandrup, Immergut, Grulke, Abe, and Bloch]{brandrup1999polymer}
Brandrup,~J.; Immergut,~E.~H.; Grulke,~E.~A.; Abe,~A.; Bloch,~D.~R. \emph{Polymer handbook}; Wiley New York, 1999; Vol.~89\relax
\mciteBstWouldAddEndPuncttrue
\mciteSetBstMidEndSepPunct{\mcitedefaultmidpunct}
{\mcitedefaultendpunct}{\mcitedefaultseppunct}\relax
\EndOfBibitem
\bibitem[Thompson \latin{et~al.}(2022)Thompson, Aktulga, Berger, Bolintineanu, Brown, Crozier, in't Veld, Kohlmeyer, Moore, Nguyen, \latin{et~al.} others]{thompson2022lammps}
Thompson,~A.~P.; Aktulga,~H.~M.; Berger,~R.; Bolintineanu,~D.~S.; Brown,~W.~M.; Crozier,~P.~S.; in't Veld,~P.~J.; Kohlmeyer,~A.; Moore,~S.~G.; Nguyen,~T.~D.; others LAMMPS-a flexible simulation tool for particle-based materials modeling at the atomic, meso, and continuum scales. \emph{Computer Physics Communications} \textbf{2022}, \emph{271}, 108171\relax
\mciteBstWouldAddEndPuncttrue
\mciteSetBstMidEndSepPunct{\mcitedefaultmidpunct}
{\mcitedefaultendpunct}{\mcitedefaultseppunct}\relax
\EndOfBibitem
\bibitem[Wang \latin{et~al.}(2004)Wang, Wolf, Caldwell, Kollman, and Case]{wang2004development}
Wang,~J.; Wolf,~R.~M.; Caldwell,~J.~W.; Kollman,~P.~A.; Case,~D.~A. Development and testing of a general amber force field. \emph{Journal of Computational Chemistry} \textbf{2004}, \emph{25}, 1157--1174\relax
\mciteBstWouldAddEndPuncttrue
\mciteSetBstMidEndSepPunct{\mcitedefaultmidpunct}
{\mcitedefaultendpunct}{\mcitedefaultseppunct}\relax
\EndOfBibitem
\bibitem[Potoff and Siepmann(2001)Potoff, and Siepmann]{potoff2001vapor}
Potoff,~J.~J.; Siepmann,~J.~I. Vapor--liquid equilibria of mixtures containing alkanes, carbon dioxide, and nitrogen. \emph{AIChE Journal} \textbf{2001}, \emph{47}, 1676--1682\relax
\mciteBstWouldAddEndPuncttrue
\mciteSetBstMidEndSepPunct{\mcitedefaultmidpunct}
{\mcitedefaultendpunct}{\mcitedefaultseppunct}\relax
\EndOfBibitem
\bibitem[Frenkel and Smit(2023)Frenkel, and Smit]{frenkel2023understanding}
Frenkel,~D.; Smit,~B. \emph{Understanding molecular simulation: from algorithms to applications}; Elsevier, 2023\relax
\mciteBstWouldAddEndPuncttrue
\mciteSetBstMidEndSepPunct{\mcitedefaultmidpunct}
{\mcitedefaultendpunct}{\mcitedefaultseppunct}\relax
\EndOfBibitem
\bibitem[Khawaja \latin{et~al.}(2017)Khawaja, Sutton, and Mostofi]{khawaja2017molecular}
Khawaja,~M.; Sutton,~A.; Mostofi,~A. Molecular simulation of gas solubility in nitrile butadiene rubber. \emph{The Journal of Physical Chemistry B} \textbf{2017}, \emph{121}, 287--297\relax
\mciteBstWouldAddEndPuncttrue
\mciteSetBstMidEndSepPunct{\mcitedefaultmidpunct}
{\mcitedefaultendpunct}{\mcitedefaultseppunct}\relax
\EndOfBibitem
\bibitem[LAM()]{LAMMPSwidom}
\url{https://docs.lammps.org/fix_widom.html}\relax
\mciteBstWouldAddEndPuncttrue
\mciteSetBstMidEndSepPunct{\mcitedefaultmidpunct}
{\mcitedefaultendpunct}{\mcitedefaultseppunct}\relax
\EndOfBibitem
\bibitem[Longuet-Higgins and Widom(1964)Longuet-Higgins, and Widom]{longuet1964rigid}
Longuet-Higgins,~H.; Widom,~B. A rigid sphere model for the melting of argon. \emph{Molecular Physics} \textbf{1964}, \emph{8}, 549--556\relax
\mciteBstWouldAddEndPuncttrue
\mciteSetBstMidEndSepPunct{\mcitedefaultmidpunct}
{\mcitedefaultendpunct}{\mcitedefaultseppunct}\relax
\EndOfBibitem
\bibitem[Mocak \latin{et~al.}(1997)Mocak, Bond, Mitchell, and Scollary]{mocak1997statistical}
Mocak,~J.; Bond,~A.~M.; Mitchell,~S.; Scollary,~G. A statistical overview of standard (IUPAC and ACS) and new procedures for determining the limits of detection and quantification: application to voltammetric and stripping techniques (technical report). \emph{Pure and Applied Chemistry} \textbf{1997}, \emph{69}, 297--328\relax
\mciteBstWouldAddEndPuncttrue
\mciteSetBstMidEndSepPunct{\mcitedefaultmidpunct}
{\mcitedefaultendpunct}{\mcitedefaultseppunct}\relax
\EndOfBibitem
\bibitem[Baba \latin{et~al.}(2022)Baba, Urano, Nagai, and Okazaki]{baba2022prediction}
Baba,~H.; Urano,~R.; Nagai,~T.; Okazaki,~S. Prediction of self-diffusion coefficients of chemically diverse pure liquids by all-atom molecular dynamics simulations. \emph{Journal of Computational Chemistry} \textbf{2022}, \emph{43}, 1892--1900\relax
\mciteBstWouldAddEndPuncttrue
\mciteSetBstMidEndSepPunct{\mcitedefaultmidpunct}
{\mcitedefaultendpunct}{\mcitedefaultseppunct}\relax
\EndOfBibitem
\bibitem[Wang and Hou(2011)Wang, and Hou]{wang2011application}
Wang,~J.; Hou,~T. Application of molecular dynamics simulations in molecular property prediction II: Diffusion coefficient. \emph{Journal of Computational Chemistry} \textbf{2011}, \emph{32}, 3505--3519\relax
\mciteBstWouldAddEndPuncttrue
\mciteSetBstMidEndSepPunct{\mcitedefaultmidpunct}
{\mcitedefaultendpunct}{\mcitedefaultseppunct}\relax
\EndOfBibitem
\bibitem[Robeson(1991)]{robeson1991correlation}
Robeson,~L.~M. Correlation of separation factor versus permeability for polymeric membranes. \emph{Journal of Membrane Science} \textbf{1991}, \emph{62}, 165--185\relax
\mciteBstWouldAddEndPuncttrue
\mciteSetBstMidEndSepPunct{\mcitedefaultmidpunct}
{\mcitedefaultendpunct}{\mcitedefaultseppunct}\relax
\EndOfBibitem
\bibitem[Robeson(2008)]{robeson2008upper}
Robeson,~L.~M. The upper bound revisited. \emph{Journal of Membrane Science} \textbf{2008}, \emph{320}, 390--400\relax
\mciteBstWouldAddEndPuncttrue
\mciteSetBstMidEndSepPunct{\mcitedefaultmidpunct}
{\mcitedefaultendpunct}{\mcitedefaultseppunct}\relax
\EndOfBibitem
\bibitem[Comesa{\~n}a-G{\'a}ndara \latin{et~al.}(2019)Comesa{\~n}a-G{\'a}ndara, Chen, Bezzu, Carta, Rose, Ferrari, Esposito, Fuoco, Jansen, and McKeown]{comesana2019redefining}
Comesa{\~n}a-G{\'a}ndara,~B.; Chen,~J.; Bezzu,~C.~G.; Carta,~M.; Rose,~I.; Ferrari,~M.-C.; Esposito,~E.; Fuoco,~A.; Jansen,~J.~C.; McKeown,~N.~B. Redefining the Robeson upper bounds for $CO_{2}$/$CH_{4}$ and $CO_{2}$/$N_{2}$ separations using a series of ultrapermeable benzotriptycene-based polymers of intrinsic microporosity. \emph{Energy \& Environmental Science} \textbf{2019}, \emph{12}, 2733--2740\relax
\mciteBstWouldAddEndPuncttrue
\mciteSetBstMidEndSepPunct{\mcitedefaultmidpunct}
{\mcitedefaultendpunct}{\mcitedefaultseppunct}\relax
\EndOfBibitem
\bibitem[Weinkauf and Paul(1990)Weinkauf, and Paul]{weinkauf1990effects}
Weinkauf,~D.; Paul,~D. \emph{Effects of structural order on barrier properties}; ACS Publications, 1990\relax
\mciteBstWouldAddEndPuncttrue
\mciteSetBstMidEndSepPunct{\mcitedefaultmidpunct}
{\mcitedefaultendpunct}{\mcitedefaultseppunct}\relax
\EndOfBibitem
\bibitem[Michaels and Bixler(1961)Michaels, and Bixler]{michaels1961solubility}
Michaels,~A.~S.; Bixler,~H.~J. Solubility of gases in polyethylene. \emph{Journal of Polymer Science} \textbf{1961}, \emph{50}, 393--412\relax
\mciteBstWouldAddEndPuncttrue
\mciteSetBstMidEndSepPunct{\mcitedefaultmidpunct}
{\mcitedefaultendpunct}{\mcitedefaultseppunct}\relax
\EndOfBibitem
\bibitem[Shen \latin{et~al.}(2018)Shen, Brown, and Hall]{shen2018diffusion}
Shen,~K.-H.; Brown,~J.~R.; Hall,~L.~M. Diffusion in lamellae, cylinders, and double gyroid block copolymer nanostructures. \emph{ACS Macro Letters} \textbf{2018}, \emph{7}, 1092--1098\relax
\mciteBstWouldAddEndPuncttrue
\mciteSetBstMidEndSepPunct{\mcitedefaultmidpunct}
{\mcitedefaultendpunct}{\mcitedefaultseppunct}\relax
\EndOfBibitem
\bibitem[Shen and Hall(2020)Shen, and Hall]{shen2020effects}
Shen,~K.-H.; Hall,~L.~M. Effects of ion size and dielectric constant on ion transport and transference number in polymer electrolytes. \emph{Macromolecules} \textbf{2020}, \emph{53}, 10086--10096\relax
\mciteBstWouldAddEndPuncttrue
\mciteSetBstMidEndSepPunct{\mcitedefaultmidpunct}
{\mcitedefaultendpunct}{\mcitedefaultseppunct}\relax
\EndOfBibitem
\bibitem[Shen and Hall(2020)Shen, and Hall]{shen2020ion}
Shen,~K.-H.; Hall,~L.~M. Ion conductivity and correlations in model salt-doped polymers: Effects of interaction strength and concentration. \emph{Macromolecules} \textbf{2020}, \emph{53}, 3655--3668\relax
\mciteBstWouldAddEndPuncttrue
\mciteSetBstMidEndSepPunct{\mcitedefaultmidpunct}
{\mcitedefaultendpunct}{\mcitedefaultseppunct}\relax
\EndOfBibitem
\bibitem[Gurnani(2021)]{gurnani2022debugging}
Gurnani,~R.~P. Debugging Neural Networks. 2021; \url{https://nanohub.org/resources/netdebugger}\relax
\mciteBstWouldAddEndPuncttrue
\mciteSetBstMidEndSepPunct{\mcitedefaultmidpunct}
{\mcitedefaultendpunct}{\mcitedefaultseppunct}\relax
\EndOfBibitem
\bibitem[Head \latin{et~al.}(2018)Head, MechCoder, Shcherbatyi, \latin{et~al.} others]{head2018scikit}
Head,~T.; MechCoder,~G.~L.; Shcherbatyi,~I.; others scikit-optimize/scikit-optimize: v0. 5.2. \emph{Version v0} \textbf{2018}, \emph{5}\relax
\mciteBstWouldAddEndPuncttrue
\mciteSetBstMidEndSepPunct{\mcitedefaultmidpunct}
{\mcitedefaultendpunct}{\mcitedefaultseppunct}\relax
\EndOfBibitem
\bibitem[PACE(2017)]{PACE}
PACE {P}artnership for an {A}dvanced {C}omputing {E}nvironment ({PACE}). 2017\relax
\mciteBstWouldAddEndPuncttrue
\mciteSetBstMidEndSepPunct{\mcitedefaultmidpunct}
{\mcitedefaultendpunct}{\mcitedefaultseppunct}\relax
\EndOfBibitem
\end{mcitethebibliography}
\end{document}

\maketitle

\newpage

\begin{figure}[H]
    \includegraphics[width=.45\columnwidth]{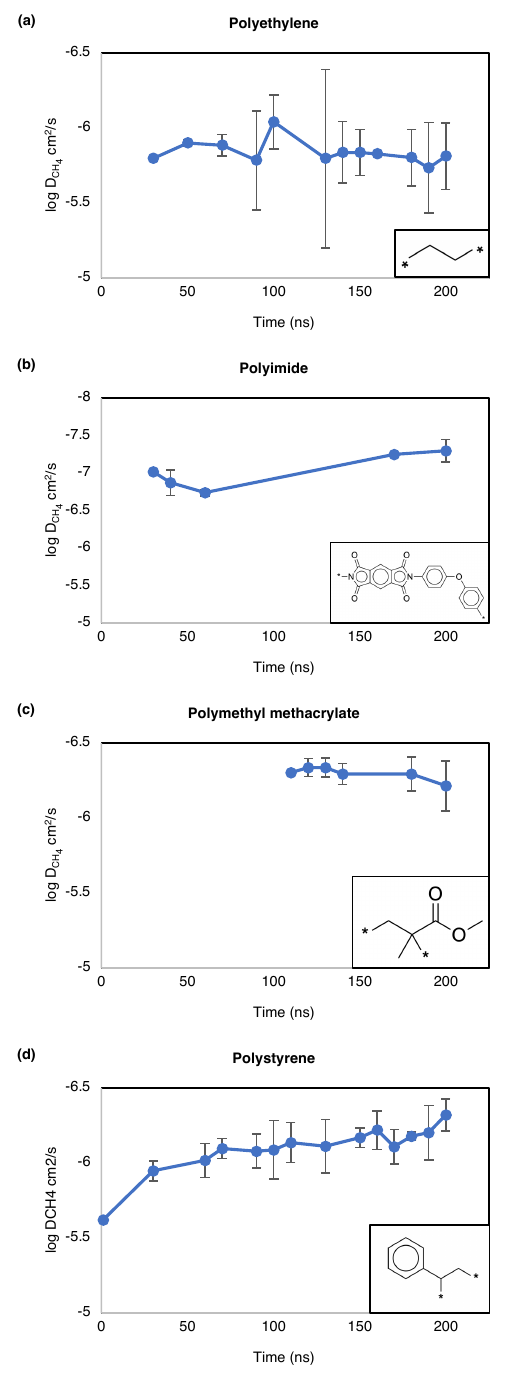}
\caption{\textbf{Simulation time vs methane diffusivity for (a) Polyethylene, (b) Polyimide, (c) Polystyrene, (d) Polymethyl methacrylate}. The log-log slopes of all the systems’ MSD curves are in the range of 0.95-1.05. Our investigations revealed that running simulations in the range of 100\textendash200$ns$ tends to yield stable diffusivity results. Polymer structures are also illustrated.
\label{fig:results}}
\end{figure}

\begin{table}
    \centering
    \begin{tabular}{|c|c|} \hline 
         Model parameters& Values\\ \hline 
         RANDOM\_SEED& 100\\ \hline 
         HP\_EPOCHS& 200\\ \hline 
         SUBMODEL\_EPOCHS& 1000\\ \hline 
         N\_FOLDS& 5\\ \hline 
         HP\_NCALLS& 25\\ \hline 
         MAX\_BATCH\_SIZE& 300\\ \hline 
         capacity\_ls& 2 to 14\\ \hline 
         weight\_decay& 0\\ \hline 
         N\_PASSES& 10\\ \hline
    \end{tabular}
    \caption{Parameters used to train benchmark and production polyGNN models.\color{black}}
    \label{tab:PolyGNN model parameters}
\end{table}

\begin{figure}[H]
    \includegraphics[width=.5\columnwidth]{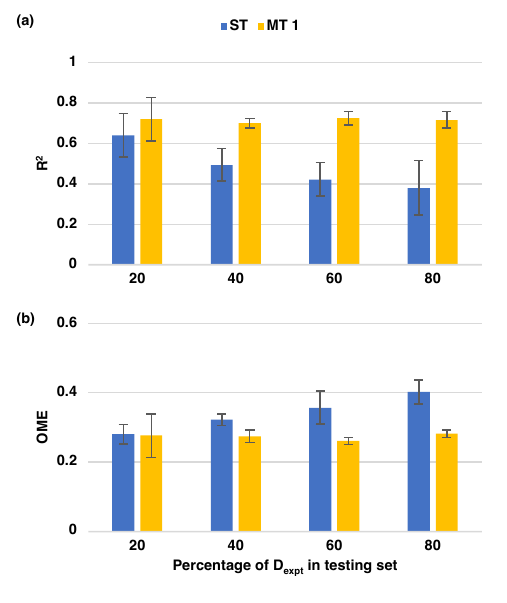}
\caption{\textbf{Predictions of Diffusivity at various train test splits (a) Coefficient of determination (R$^2$) and (b) order of magnitude error (OME)}, comparisons between the ST and MT models as a function of percentage (\%) of the unseen test set. The different unseen test set sizes demonstrate the effect of having less and less data in the training set. At 80\%, the model only sees 20\% of the dataset and is tested on the remaining 80\%. This 80\% replicates a data-scarce region, where there is less chemical coverage. A comprehensive improvement is seen in MT over ST models, especially at higher unseen test set \%. \label{fig:results}}
\end{figure}

\begin{figure}[H]
    \includegraphics[width=.5\columnwidth]{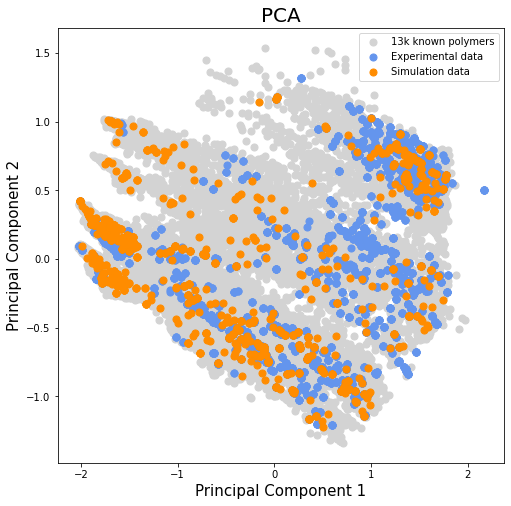}
    \caption{\textbf{Principal Component Analysis (PCA) plot.} The PCA plot demonstrates the chemical coverage of our simulation data in comparison to experimental gas transport data and 13,000 known polymers in our database. The blue and orange dots corresponds to experimental and simulation data respectively. The grey dots represent the 13,000 known polymers in our database.\color{black}
\label{prod_pca}}
\end{figure}

\begin{figure}[H]
    \includegraphics[width=.5\columnwidth]{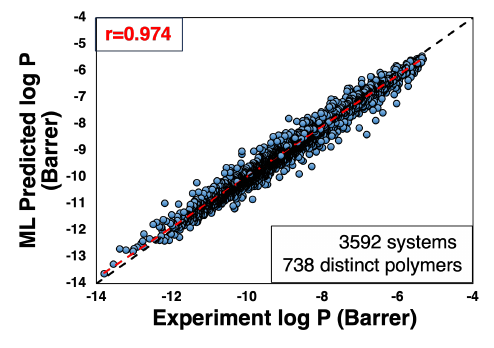}
    \caption{\textbf{Train Parity Plot.} Comparison of predicted and experimental permeability values in train set using the MT production model.
\label{fig:train}}
\end{figure}

\begin{figure}[H]
    \includegraphics[width=.5\columnwidth]{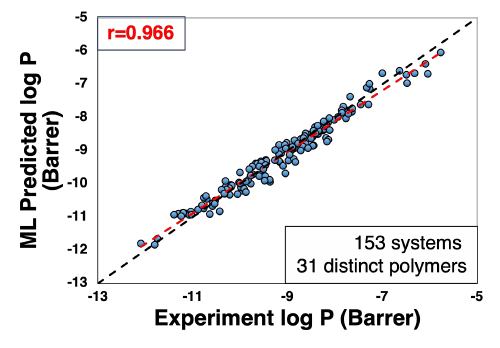}
    \caption{\textbf{Test Parity Plot.} Comparison of predicted and experimental permeability values in test set using the MT production model.
\label{fig:test}}
\end{figure}

\begin{table}[]
\centering
\resizebox{\textwidth}{!}{%
\begin{tabular}{|
>{\columncolor[HTML]{FFFFFF}}c |
>{\columncolor[HTML]{FFFFFF}}c |
>{\columncolor[HTML]{FFFFFF}}c |
>{\columncolor[HTML]{FFFFFF}}l |}
\hline
Polymer Class                      & Polymer Name                                                                  & Datapoints & R$^{2}$   \\ \hline
Conjugated Polymers                & Poly(phenylene vinylene)                                                      & 4          & 0.97 \\ \hline
Polycarbonates                     & Hexafluoro Bis A polycarbonate                                                & 6          & 0.99 \\ \hline
Polyarylates                       & \begin{tabular}[c]{@{}c@{}}TBHFBPA-I PAR\\ TMBPA-tBI polyarylate\end{tabular} & 12         & 0.97 \\ \hline
Polynorbornenes                    & PB3FMPNDI                                                                     & 5          & 0.95 \\ \hline
Polyimides and Polypyrrolones &
  \begin{tabular}[c]{@{}c@{}}6FDA-6FpDA PI \\ TADATO-TMBZ PI \\ 6FDA-4BDAF \\ BTDA-6FpDA \\ PMDA-3,3'-ODA PI \\ DSDA-DDBT PI \\ BPDA-BAPP PI \\ 6FDA-DMMDA PI \\ 6FDA/PPTI-4 PI \\ HQDPA-TMMDA1 PEI \\ BPI-BTDA PI\end{tabular} &
  53 &
  0.99 \\ \hline
Vinyl and Vinylidene Polymers &
  \begin{tabular}[c]{@{}c@{}}Poly(vinyl acetate)\\ Poly(p-trimethylsilyl styrene)\end{tabular} &
  9 &
  0.99 \\ \hline
Poly(aryl ethers) and Poly(aryl ether ketones) &
  \begin{tabular}[c]{@{}c@{}}6H-PEN poly(aryl ether)\\ Poly(aryl ether ketone) TMPEK-C\end{tabular} &
  10 &
  0.98 \\ \hline
Polyphosphazenes                   & Polyphosphazene 4                                                             & 5          & 0.86 \\ \hline
Polysulfones                       & DMPSF                                                                         & 5          & 0.99 \\ \hline
Polyamides and poly(amide- imides) & APA1 aromatic polyamide                                                       & 5          & 0.99 \\ \hline
Parylenes                          & Poly(methyl-p-xylylene)                                                       & 4          & 0.97 \\ \hline
High Temperature Polymers          & \begin{tabular}[c]{@{}c@{}}PIM-PI-7\\ Poly(1,2,4-triazole)\end{tabular}       & 10         & 0.99 \\ \hline
Polypropynes, substituted polyacetylenes, polypentynes &
  \begin{tabular}[c]{@{}c@{}}Polyacetylene \\ Poly{[}1-(n-hexyllthio)-1-propyne{]} \\ Fluorenyl-poly(diarylacetylene)\\ Poly(diphenyl acetylene)Poly(2-nonyne)\end{tabular} &
  26 &
  0.94 \\ \hline
\end{tabular}%
\caption{An overview of the polymer classes, the number of data points, and the performance metrics used in the comparison of the original and new model.}
}
\end{table}

\begin{figure}[H]
    \includegraphics[width=1\columnwidth]{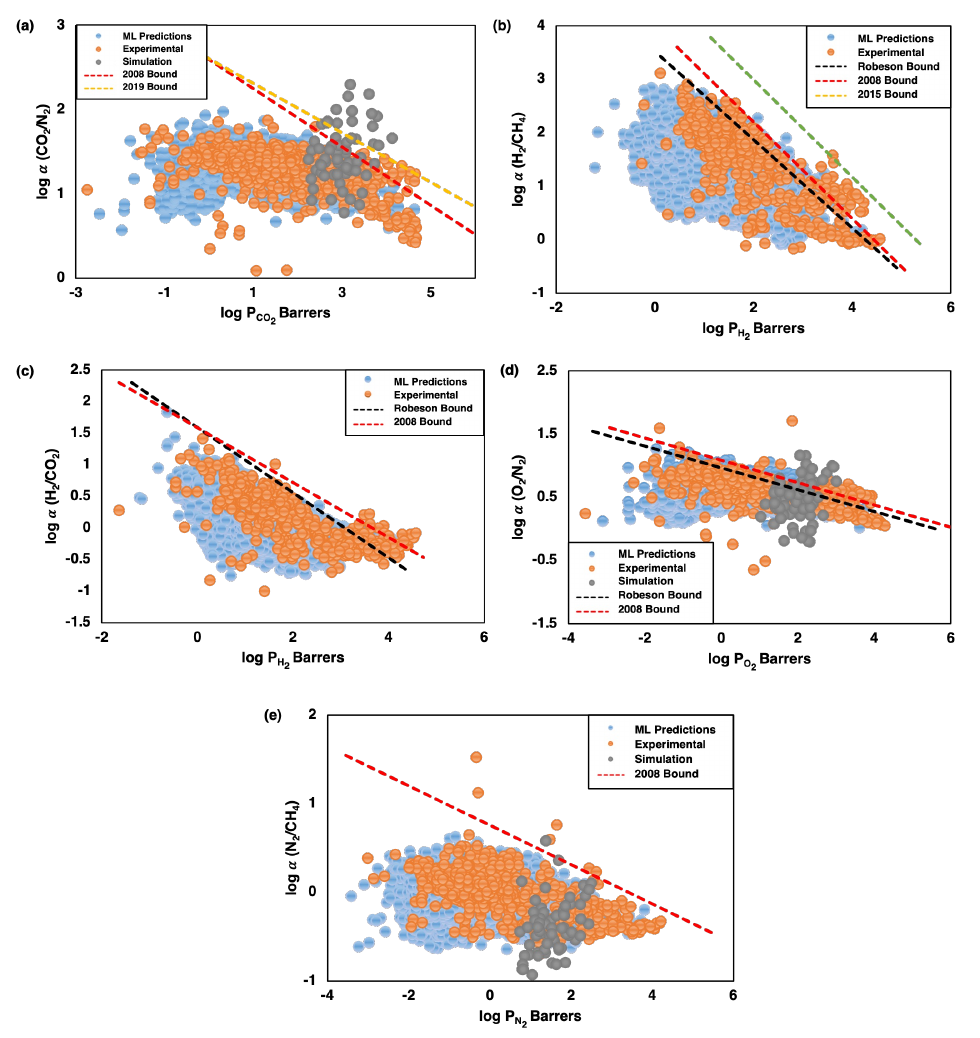}
    \caption{\textbf{Gas permeability trade-off plots for (a) CO$_{2}$/N$_{2}$, (b) H$_{2}$/CH$_{4}$, (c) H$_{2}$/CO$_{2}$, (d) O$_{2}$/N$_{2}$ and (e) N$_{2}$/CH$_{4}$}
    Polymer upper bound with predicted and experimental gas permeability and selectivity data points.
\label{fig:p_tradeoff}}
\end{figure}

\begin{figure}[H]
    \includegraphics[width=1\columnwidth]{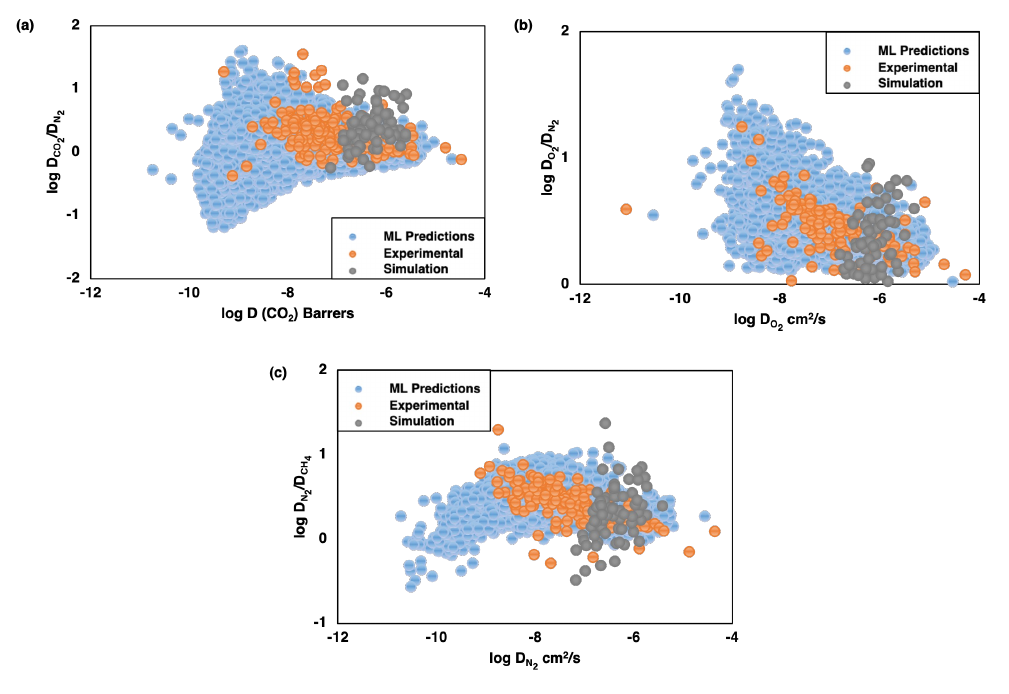}
    \caption{\textbf{Gas diffusivity trade-off plots for (a) CO$_{2}$/N$_{2}$, (b) O$_{2}$/N$_{2}$ and (c) N$_{2}$/CH$_{4}$. }Polymer trade-off plots with predicted and experimental gas diffusivity and selectivity data points.
\label{fig:d_tradeoff}}
\end{figure}

\begin{figure}[H]
    \includegraphics[width=1\columnwidth]{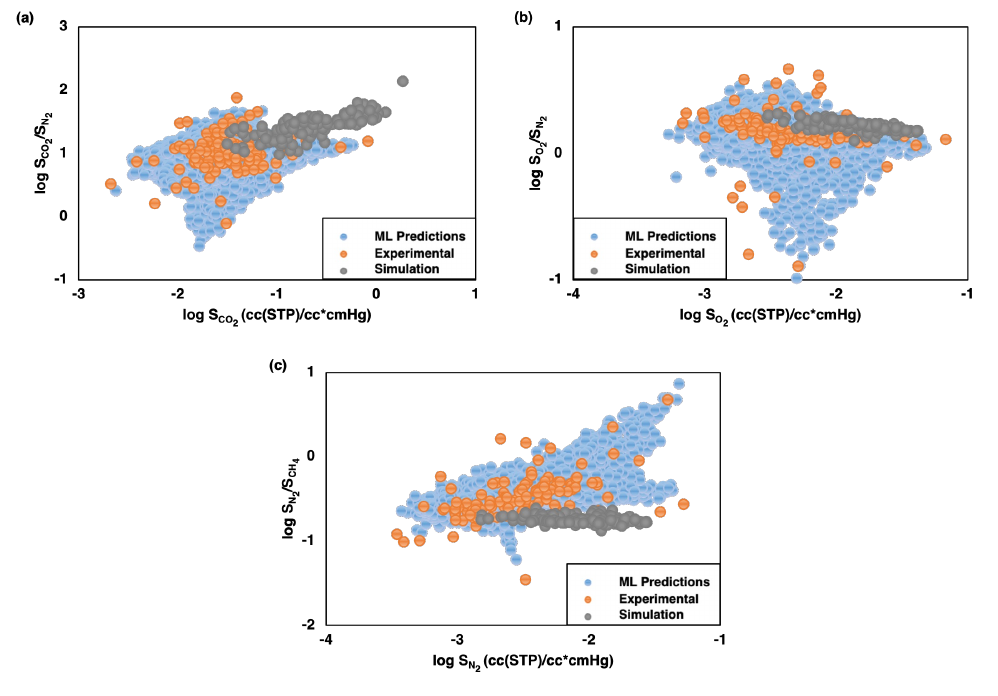}
    \caption{\textbf{Gas solubility trade-off plots for (a) CO$_{2}$/N$_{2}$, (b) O$_{2}$/N$_{2}$ and (c) N$_{2}$/CH$_{4}$.} Polymer trade-off plots with predicted and experimental gas solubility and selectivity data points.
\label{fig:s_tradeoff}}
\end{figure}